\documentclass[10pt, twocolumn, twoside]{article}
\usepackage{framed,multirow}
\usepackage[round]{natbib}
\usepackage{graphicx}
\usepackage{amssymb}
\usepackage{latexsym}
\usepackage[top=3cm,bottom=3cm, left=2.2cm, right=2.2cm]{geometry}
\usepackage{url}
\usepackage{xcolor}

\usepackage{hyperref}
\usepackage[utf8]{inputenc}
\usepackage[T1]{fontenc}
\usepackage{graphicx}
\usepackage{booktabs}
\usepackage{enumitem}
\usepackage{cite}
\usepackage{capt-of}
\usepackage{makecell}
\usepackage{graphbox}
\usepackage{ragged2e}
\usepackage{placeins}
\usepackage{fancyhdr}
\usepackage{amsmath,amsfonts,bm}

\newcolumntype{P}[1]{>{\RaggedRight\hspace{0pt}}p{#1}}

\definecolor{newcolor}{rgb}{.8,.349,.1}
\newcommand\review[1]{{\color{black}#1}}

\begin{document}

\title{\textbf{FetMRQC: a robust quality control system\\ for multi-centric fetal brain MRI}}%
\author{Thomas Sanchez$^{1,2,*}$, Oscar Esteban$^{2}$, Yvan Gomez$^{3,4}$, Alexandre Pron$^{5}$,\\ Mériam Koob$^{2}$, Vincent Dunet$^{2}$, Nadine Girard$^{5,6}$, Andras Jakab$^{7,8,9}$,\\ Elisenda Eixarch$^{3,10}$, Guillaume Auzias$^{5}$, Mertixell Bach Cuadra$^{1,2}$\\
\normalsize$^*$\texttt{thomas.sanchez@unil.ch}\\[2mm]
\small$^1$CIBM -- Center for Biomedical Imaging, Switzerland\\[-1mm]
\small$^2$Department of Diagnostic and Interventional Radiology, Lausanne University Hospital\\[-1mm] \small and University of Lausanne, Lausanne, Switzerland \\[-1mm]
\small$^3$BCNatal Fetal Medicine Research Center (Hospital Clínic and Hospital Sant Joan  de Déu),\\[-1mm] \small  Universitat de Barcelona, Spain\\[-1mm]
\small$^4$Department Woman-Mother-Child, Lausanne University Hospital, Lausanne, Switzerland\\[-1mm]
\small$^5$Aix-Marseille Université, CNRS, Institut de Neurosciences de La Timone, Marseilles, France\\[-1mm]
\small$^{6}$Service de Neuroradiologie Diagnostique et Interventionnelle, Hôpital Timone, AP-HM, Marseilles, France\\[-1mm]
\small$^7$Center for MR Research, University Children’s Hospital Zurich, University of Zurich, Zurich, Switzerland\\[-1mm]
\small$^8$Neuroscience Center Zurich, University of Zurich, Zurich, Switzerland\\[-1mm]
\small$^9$Research Priority Project Adaptive Brain Circuits in Development and Learning (AdaBD),\\[-1mm]\small University of Zürich, Zurich, Switzerland\\[-1mm]
\small$^{10}$IDIBAPS and CIBERER, Barcelona, Spain\\[-1mm]}
\date{}

\twocolumn[
\maketitle
\begin{abstract}
Fetal brain MRI is becoming an increasingly relevant complement to neurosonography for perinatal diagnosis, allowing fundamental insights into fetal brain development throughout gestation. However, uncontrolled fetal motion and heterogeneity in acquisition protocols lead to data of variable quality, potentially biasing the outcome of subsequent studies. We present FetMRQC, an open-source machine-learning framework for automated image quality assessment and quality control that is robust to domain shifts induced by the heterogeneity of clinical data. FetMRQC extracts an ensemble of quality metrics from unprocessed anatomical MRI and combines them to predict experts' ratings using random forests. We validate our framework on a pioneeringly large and diverse dataset of more than 1600 manually rated fetal brain T2-weighted images from four clinical centers and 13 different scanners. Our study shows that FetMRQC's predictions generalize well to unseen data while being interpretable. FetMRQC is a step towards more robust fetal brain neuroimaging, which has the potential to shed new insights on the developing human brain. 

\vspace{0.5cm}


\end{abstract}
\vspace{4cm}
{\footnotesize This work has been accepted for publication at Medical Image Analysis.\\The final version is available at \url{https://doi.org/10.1016/j.media.2024.103282}.}

]

\section{Introduction}
Establishing a protocol for objective image quality assessment and control for neuroimaging studies is critical to enforce reliability, generalization and replicability~\citep{mortamet2009automatic,niso2022open,rosen2018quantitative}. Quality assessment (QA)  focuses on assessing and eventually improving the quality of a process to prevent issues from propagating, while quality control (QC) looks to find and discard problematic outputs of that process~\citep{alfaro2018image}. Both steps are fundamental in magnetic resonance imaging (MRI) studies, as insufficient MRI data quality has been shown to bias statistical analyses and neuroradiological interpretation~\citep{power2012spurious,reuter2015head,alexander2016subtle}. 

\begin{figure*}[!t]
    \centering
    \includegraphics[width=.9\linewidth, trim=2 2 2 2, clip]{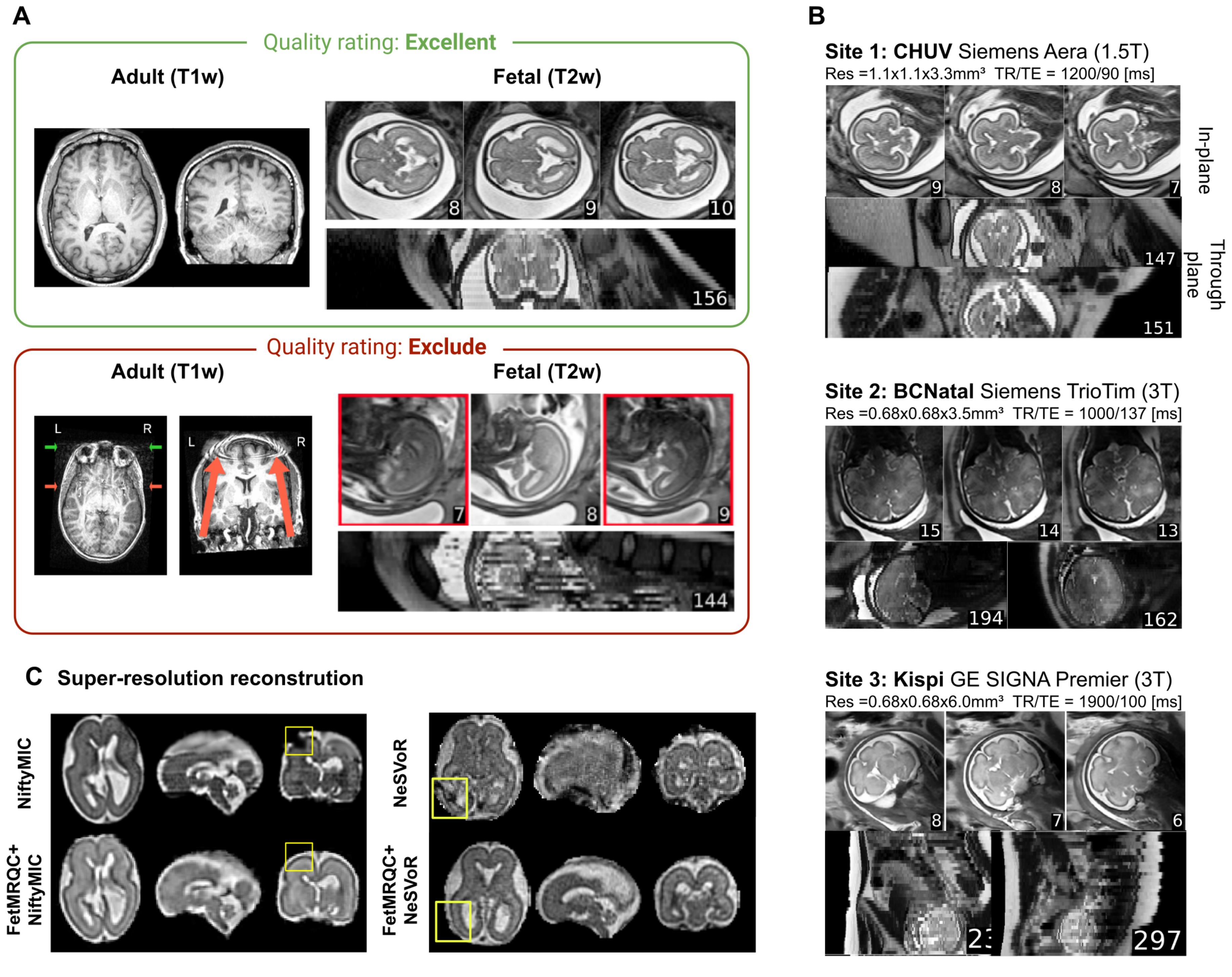}
    \caption{\textbf{Variations in data quality illustrated.} \textbf{A} -- Comparison of data across adult (T1w), from the ABIDE dataset~\citep{di2014autism} and from fetal acquisitions. In the excluded scans, the adult image on the left suffers from severe motion artifacts, while large coil artifacts corrupt the image on the right. The fetal data suffer from strong intensity changes between multiple slices and signal drop; in the through-plane view, strong inter-slice motion makes it difficult to discern the brain structures. \textbf{B}~--~Examples of data acquired on different scanners, with very different appearance. The in-plane and through-plane resolution, the field of view, the repetition time (TR), and the echo time (TE) can all substantially change between acquisition protocols. \textbf{C}~--~Importance of quality control for super-resolution reconstruction (SRR), illustrated using NiftyMIC~\citep{ebner_automated_2020}, and NeSVoR~\citep{xu2023nesvor}, two SRR methods with built-in outlier rejection. On the top row a subject is reconstructed using all stacks available (13 for NiftyMIC, 5 for NeSVoR), and each reconstruction shows large artifacts. On the bottom row, FetMRQC is plugged in and by removing low quality series (6 out of 13 for NiftyMIC, 2 out of 5 for NeSVoR), the reconstruction quality is improved.}
    \label{fig:qc_example}
\end{figure*}

Automated QA/QC tools designed to assist data exclusion decisions for adult brain neuroimaging studies ~\citep{esteban2017mriqc,klapwijk2019qoala,vogelbacher2019lab,ravi2023efficient} are becoming increasingly available. However, these techniques are inapplicable to fetal MRI, as they rely on priors that are not valid \textit{in utero}, such as e.g., assuming that the head is surrounded by air or the relative orientation of the brain with respect to the stereotaxic frame defined by the scanner. In addition, fetal brain MRI typically displays larger and uncontrolled motion of the head as fixation techniques (e.g., padding) and real-time feedback countermeasures are only available after birth (Fig. \ref{fig:qc_example}A). Moreover, fetal brain imaging greatly lacks standardization in acquisition protocols (Fig. \ref{fig:qc_example}B). While consensus has settled on 2-dimensional (2D) fast-spin echo interleaved T$_\text{2}$-weighted (T2w) MR schemes showcasing thick slices~\citep{tortori2005fetal,gholipour2014fetal}, specific imaging parameters such as in-plane resolution, slice thickness, field of view, or vendor implementation of the imaging sequence greatly vary.
As a result, the appearance and quality of fetal MR images in this wild-type data vary markedly across centers (Fig. \ref{fig:qc_example}B).

Although fetal brain MRI can be severely affected by artifacts like inter-slice motion, signal drops or bias field~\citep{gholipour2014fetal}, only few methods dedicated to QA/QC have been proposed. 
Initially, automated QA/QC has been integrated within the super-resolution reconstruction (SRR) process~\citep{uus2022automated,kuklisova-murgasova_reconstruction_2012,ebner_automated_2020,tourbier_efficient_2015,xu2023nesvor}.  
SRR is a ubiquitous early step of the fetal MRI processing workflow that builds a high-resolution, isotropic, 3D volume from several differently-oriented stacks of 2D slices with low-resolution (LR) along the through-plane axis (i.e., anisotropic resolution)~\citep{uus2022retrospective}.
Some of the proposed approaches incorporate an automated QC stage for outlier rejection that excludes sub-standard slices or pixels from the input low-resolution stacks, and measure the similarity between a reconstructed slice and an input slice using information-theoretic metrics~\citep{ebner_automated_2020,kuklisova-murgasova_reconstruction_2012,kainz_fast_2015,xu2023nesvor}. However, as illustrated on Fig. \ref{fig:qc_example}c, sub-optimal quality stacks can remain detrimental to the final quality of the reconstruction, even when SRR pipelines include outlier rejection schemes. Additional QA/QC checkpoints are thus needed to filter out low-quality raw T2 stacks before using SRR, and several deep learning-based methods were recently proposed for this task~\citep{lala2019deep,xu2020semi,liao2020joint}. These solutions aim to automatically identify problematic slices for exclusion (QC), and, if streamlined with the acquisition, enable re-acquiring corrupted slices on the fly~\citep{gagoski2022automated} (QA). However, these methods operate at the slice level, and not all artifacts can be seen by analyzing slices independently. For instance, inter-slice motion (visible on the right of Figure 1a), a strong bias field in the through-plane direction, or an incomplete field of view can be spotted only when considering the entire stack of slices. Stack-wise QA/QC methods are thus still needed.

Importantly, these methods face the challenge of deployment to unseen scanners or acquisition settings: how will they generalize to unseen domains? Due to the private and sensitive nature of medical data~\citep{willemink2020preparing}, building large and diverse medical imaging datasets is difficult endeavor.
As a consequence, proposed methods are often only evaluated on locally available data, and can fail to deal with the heterogeneity found across different centers~\citep{sambasivan2021everyone,varoquaux2022machine}. 
In addition, while openly shared MRI databases have been released for adults~\citep{mueller2005ways,di2014autism,markiewicz2021openneuro,van2013wu}, children and adolescents~\citep{makropoulos2018developing,casey2018adolescent}, privacy protection regulations and ethical limitations to data-sharing are much stronger regarding fetuses, making it even more difficult to construct robust ML models trained on multicentric data. As today, the question of the robustness of state-of-the-art approaches to fetal brain quality control~\citep{xu2020semi,ebner_automated_2020,uus2022automated} to unseen domains remains open. 

Beyond the need of supporting SRR, quality assessment also builds towards reproducible neuroimaging pipelines, allowing to fairly compare different processing steps~\citep{payette_automatic_2021}. For instance, initiative of fetal brain tissue segmentation but lack of systematic/standardized objective evaluation of quality input data that would support the analysis of the comparison results~\citep{payette2023fetal}.  

The contribution of our paper is threefold. First, we introduce  a framework specifically designed for QA/QC manual annotations of T2w fetal brain MRI. It generates a visual report for efficient stack screening and manual QA, facilitating the work of raters. Second, we present FetMRQC, a machine learning model based on manual ratings to automatically perform two tasks: 1) quality assessment, where a discrete score between 0 (bad quality) to 4 (excellent) quality is predicted, and 2) quality control, where the model predicts whether an image reaches a predefined quality threshold. QA~-- a regression task in our case -- and QC -- a binary classification problem --  are performed automatically by a random forest that uses an ensemble of 332 image quality metrics (IQMs), extracted from raw T2w stacks, that reflect complementary quality features based on various statistics computed from image intensity, brain mask and segmentation (details on IQMs extraction is available in the Materials and Methods section).
Third, by collecting and manually annotating a very large collection of 1649 low-resolution T2w images from 233 subjects, acquired in 13 different scanners in four different institutions across Europe, we can benchmark the generalization of automated QA/QC models to unseen domains, including existing baselines~\citep{ebner_automated_2020} and pre-trained deep learning models~\citep{legorreta2020DL,xu2020semi}. A pilot study of this work, including fewer IQMs and only two centers, was previously presented~\citep{sanchez2023fetmrqc}. The code and image quality metrics are available at \url{https://github.com/Medical-Image-Analysis-Laboratory/fetmrqc}.


\section{Methods}

\subsection{Data}
For this study, we retrieved 1649 T2-weighted 2D stacks of slices from 233 subjects from existing databases at four different institutions, including both neurotypical and pathological cases. The corresponding local ethics committees independently approved the studies under which data were collected, and all participants gave written informed consent. 

Lausanne University Hospital (CHUV), Switzerland, provided 61 subjects (498 scans), with an average of 7.9$\pm$3.0 stacks per subject. BCNatal (Hospital Sant Joan de Déu, Barcelona, Spain) provided 85 subjects (508 scans), $5.8\pm3.4$ stacks per subject. University Children's Hospital Zürich (KISPI), Switzerland, provided 19 subjects (441 scans) with 23.2$\pm$5.36 stacks per subject. La Timone University Hospital, Marseille, France, provided 68 subjects (203 scans) with 3 stacks per subject. The reason for having few scans per subject at La Timone is due to the acquisition duration being limited in clinical routine, while other centers have a more research-oriented acquisition. After the exclusion of scanners with insufficient data (CHUV - Siemens Avanto with 5 stacks), the aggregate sample size is N=1644 stacks. The imaging parameters, magnetic field strength, repetition time (TR), echo time (TE), field of view (FoV), etc. greatly varied across centers and scanners, reflecting the heterogeneity found in clinical practice. The details are provided in Table \ref{tab:scanners}.

The acquisition parameters show a very large variability across scanners and sites. For instance, the resolution of 1.5~T scanners changes from $1.1\times1.1$ $\text{mm}^2$ (e.g. CHUV - Aera) in-plane to 0.5$\times$0.5 $\text{mm}^2$ (e.g. KISPI - Signa Artist), which leads to large differences in signal-to-noise ratio. In addition, different models using the similar parameters can also yield largely different images. Examples are shown on Figure \ref{fig:qc_example}B. Such variable parameters are strong indicators of domain shifts that might challenge the generalization of machine learning models.

\begin{table*}[!t]
\centering
    \caption{Detailed description of the data used in the \review{study}. Field refers to the magnetic field of the scanner, TR is the repetition time and TE is the echo time, FoV is the field of view.}\label{tab:scanners}
    \resizebox{0.7\linewidth}{!}{\begin{tabular}{lclcccc}
    \toprule
    \multicolumn{6}{c}{\textsc{\textbf{CHUV}}}\\
    \textbf{Model (Siemens)} & Field [T] & $(n_{\text{subjects}},n_{\text{LR}})$ & TR [ms] & TE [ms] & Resolution [$\text{mm}^3$]  & FoV [cm]\\
    \midrule
    \textbf{Aera}            &1.5    &$(34,281)$ &  1200 & 90    & $1.12\times1.12\times3.3$ & $36$\\
    \textbf{MAGNETOM Sola}   & 1.5   &$(17, 138)$& 1200  & 90    & $1.1\times1.1\times3.3$ & $36$\\
    \textbf{MAGNETOM Vida}~~~& 3     &$(2, 14)$  & 1100  & 101   & $0.55\times0.55\times3$ & $35$\\
    \textbf{Skyra}           &3      &$(8, 77)$  &  1100 & 90    & $0.55\times0.55\times3$ & $35$\\
    \midrule
    \multicolumn{6}{c}{\textsc{\textbf{BCNatal}}}\\
    \textbf{Model (Siemens)} & Field [T] & $(n_{\text{subjects}},n_{\text{LR}})$ & TR [ms] & TE [ms] & Resolution [$mm^3$]  & FoV [cm]\\
    \midrule
    \textbf{Aera}   & 1.5   & $\mathbf{(16, 158)}$ &&&\\
                            &       &-~$(6,80)$  & 1500& 82& $0.55\times0.55\times2.5$ & $28$\\
                            &       &-~$(4,34)$   & 1000& 137& $0.59\times0.59\times3.5$ & $23$ / $30$\\
                            &       &-~$(4, 33)$ & 1000&81 & $0.55\times0.55\times3.15$ & $28$\\
                            &       &-~$(2,11)$   & 1200& 94& $1.72\times1.72\times4.2$ & $36$ / $44$\\
    \textbf{MAGNETOM Vida}   &   3   & $(11, 56)$& 1540 & 77 & $1.04\times1.04\times3$ & $20$\\
    \textbf{TrioTim}         &3      & $\mathbf{(59, 322)}$&&&  \multicolumn{2}{r}{\textit{// 4 outliers}}\\
                            &       &-~$(24,97)$& 1100&127&$0.51\times0.51\times3.5$ & $26$\\
                            &       & -~$(15,108)$& 990& 137& $0.68\times0.68\times3.5-6.0$ &$26$\\
                            &       &-~$(14,71)$& 2009& 137& $0.51\times0.51\times3.5$ &$26$\\
                            &       &-~$(1,14)$ & 3640& 137& $0.51\times 0.51\times 3.5$ &$26$\\
    \midrule                        
    \multicolumn{6}{c}{\textsc{\textbf{Kispi}}}\\
    \textbf{Model (General Electric)} & Field [T] & $(n_{\text{subjects}},n_{\text{LR}})$ & TR [ms] & TE [ms] & Resolution [$mm^3$]  & FoV [cm]\\
    \midrule                 
    \textbf{SIGNA Premier}    & 3 &  $\mathbf{(3,58)}$        &&& \multicolumn{2}{r}{\textit{// 8 outliers}}\\
                       &   &  -~$(3,24)$&$<2500$&100/120&$0.65\times 0.65 \times 3/5$& $33$\\
                        &   &  -~$(3,26)$&$3000$&120&$0.47/0.57\times 0.47/0.57 \times 3$& $29/24$\\
    \textbf{Discovery MR750}          & 3 & $\mathbf{(5,125)}$ &&& \multicolumn{2}{r}{\textit{// 5 outliers}}\\
                            &   &  -~$(5,29)$&$<2500$&120&$0.65\times 0.65 \times 3/5$& $33$\\
                             &   & -~$(5,81)$&$3000$&120&$0.55\times 0.55 \times 3$& $28$\\
                             &   & -~$(5,10)$&$5000$&120/500&$0.53\times 0.53 \times 3/5$ & $28$\\
    \textbf{SIGNA Artist}    & 1.5 &  $\mathbf{(11,258)}$   &&& \multicolumn{2}{r}{\textit{// 22 outliers}}\\
                            &   &  -~$(11,108)$&$<2500$&100/120&$0.47/0.64\times 0.47/0.64 \times 3/5$& $24-35$\\
                            &   &  -~$(11,128)$&$3000$&120&$0.47/0.55\times 0.47/0.55 \times 3$& $26$\\
    \midrule                          
    \multicolumn{6}{c}{\textsc{\textbf{La Timone}}}\\
    \textbf{Model (Siemens)} & Field [T] & $(n_{\text{subjects}},n_{\text{LR}})$ & TR [ms] & TE [ms] & Resolution [$mm^3$]  & FoV [cm]\\
    \midrule                            
    \textbf{Skyra}          & 3 & $\mathbf{(34,101)}$ &&&\\
                            &   &-~$(31,93)$   &3200   &177 &$0.68\times 0.68 \times 3$& $26$\\
                            &   &-~$(3,8)$     &3750   &183 &$0.59\times 0.59 \times 3$& $30$\\
    \textbf{SymphonyTim}    & 1.5 &  (34,102)  &1680&137&$0.74\times 0.74 \times 3.5$& $38$\\         
    \bottomrule
    \end{tabular}
    }
\end{table*}

\subsection{Manual QA of fetal MRI stacks}
FetMRQC comprehends two major elements to implement QA/QC protocols of unprocessed (stacks of 2D slices) fetal brain MRI data.
First, the tool builds upon MRIQC's framework and generates an individual QA report for each stack to  assist and optimize screening and annotation by experts. Second, FetMRQC proposes to train machine learning models based on image quality metrics (IQMs).

Akin to MRIQC~\citep{esteban2017mriqc}, FetMRQC generates an HTML-based report adapted to the QA of fetal brains for each input stack of 2D slices (Figure \ref{fig:data}A) to help make the process of manual rating of quality standardized and efficient. The input dataset is required to comply with the Brain Imaging Data Structure (BIDS~\citep{gorgolewski2016brain}), a format widely adopted in the neuroimaging community. The reports are generated using an image with a corresponding brain mask. This mask can be extracted automatically, and in this work, we used MONAIfbs~\citep{ranzini2021monaifbs}. Each individual-stack report has a QA utility (the so-called rating widget), with which raters can fill in an overall quality score, the in-plane orientation, and the presence and grading of artifacts visible in the stack. We use an interval (as opposed to categorical) rating scale with four main quality ranges:  [0,1): exclude – [1,2): poor – [2,3): acceptable – [3,4): excellent. Interval ratings simplify statistical modeling, set lower bounds to annotation noise, and enable the inference task where a continuous quality score is assigned to input images rather than broad categories. In addition, a navigation menu allows the rater to access all reports in a centralized location, and by being able to access the next image to be rated in a single click.  Being HTML-based, the reports can be visualized on any web browser, and effectively remove any bias due to using different image visualization software.

\subsection{IQMs extraction and prediction models}
FetMRQC's QA/Qc prediction models work in two steps. An ensemble of image quality metrics are first extracted from the raw T2-weighted images and then are used as input to a classification or regression model that learns to predict the quality ratings from the IQMs.

\subsubsection{IQMs tailored to fetal brain MRI}
While tools designed for QA/QC for adult brain neuroimaging studies~\citep{esteban2017mriqc,klapwijk2019qoala} are available, they are not readily applicable to fetal brain MRI, due to priors invalid in this context. However, some IQMs can be translated to fetal brain MRI and several works have proposed developed quantities that can be used as IQMs, and we include them as features in FetMRQC. The method of \citet{kainz_fast_2015}, \texttt{rank\_error}, predicts the quality of a raw T2-weighted stack by estimating its compressibility using singular value decomposition. \citet{ebner_automated_2020} used the volume of the brain mask, \texttt{mask\_volume}, to exclude outlying stacks, and \citet{dedumast2020translating} computed its centroid to estimate inter-slice motion. We also inclode recently proposed slice-wise and stack-wise deep learning-based IQMs, \texttt{dl\_slice}~\citep{xu2020semi} and \texttt{dl\_stack }~\citep{legorreta2020DL}. We use their pre-trained models, as we want to test the off-the-shelf value of these IQMs. Note that the method of Liao et al.~\citep{liao2020joint} was not included because their code is not publicly available and we could not get in contact with the authors. \texttt{dl\_slice}~\citep{xu2020semi} predicts simultaneously whether a slice contains some brain volume, and whether this slice is of good quality. We aggregate their slice-wise score into a global score by computing $\frac{1}{n_\text{slices}} \sum_{i=1}^{n_{\text{slices}}} p_{i,\text{pass}}- p_{i,\text{fail}}$, yielding a score between -1 and 1. 

Along with these existing IQMs, we also propose additional IQMs for quality prediction that have not previously been used in the context of fetal brain MRI. They can be roughly categorized into three groups: intensity-based, mask-based, segmentation-based. In a nutshell, \textit{intensity-based} IQMs directly rely on the voxel values of the image. These include summary statistics~\citep{esteban2017mriqc} such as mean, median, and percentiles. We also repurpose metrics traditionally used for outlier rejection, such as PSNR or Normalized Cross Correlation (NCC)~\citep{kuklisova-murgasova_reconstruction_2012,kainz_fast_2015,ebner_automated_2020} to quantify the intensity difference between slices in a volume. We compute entropy~\citep{esteban2017mriqc}, estimate the level of bias using N4 bias field correction~\citep{tustison_n4itk_2010} and estimate the sharpness of the image with Laplace and Sobel filters. The second type of metrics are \textit{mask-based} and operate directly on the automatically extracted brain mask. We propose to use a morphological closing in the through-plane direction to detect inter-slice motion, as well as edge detection, to estimate the variation at the surface of the brain mask, using Laplace and Sobel filters. The third type of IQMs is \textit{segmentation-based}. While such metrics were originally proposed in the context of \textit{MRIQC}~\citep{esteban2017mriqc}, they have never been adapted to fetal brain imaging. These are segmentation-based and include region-wise summary statistics, region-wise volume, region-wise signal-to-noise ratio~\citep{dietrich2007measurement}, contrast-to-noise ratio between white matter (WM) and gray matter (GM)~\citep{magnotta2006measurement}, coefficient of joint variation between gray matter and white matter~\citep{ganzetti2016intensity} and white matter to maximum intensity ratio~\citep{esteban2017mriqc}. In order to compute these segmentations from the raw T2-weighted stacks, we train a nnUNet-v2~\citep{isensee2021nnu} 2D model on the FeTA dataset~\citep{payette_automatic_2021}, a public dataset consisting of super-resolution (SR) reconstructed fetal brain images along with manual segmentations. The model is trained with the parameters automatically defined by nnUNet, which yield satisfactory results for SR volumes, and is then used to perform slice-wise inference on the low-resolution T2-weighted stacks. The segmentations are done over eight different classes, which we merge then into three groups: white matter (excluding corpus callosum), cerebrospinal fluid (CSF; intra-axial and extra-axial), and gray matter (cortical and deep). This is done to enable the use of the segmentation-based IQMs from MRIQC~\citep{esteban2017mriqc}, which rely on these three groups.

\paragraph{Variants of the metrics}
All the IQMs operate by default on raw T2-weighted 2D images and/or masks, but they can be pre-processed in various manners. For example, \citet{kainz_fast_2015} evaluated their metrics only on the third of the slices closest to the center of a given volume. We construct variants on our IQMs using various pre-processing methods. The variants include considering the third of the center-most slices instead of the whole ROI; masking the maternal tissue in the background; aggregating point estimates using mean, median, or other estimators; and computing information theoretic metrics on the union or intersection of masks. Finally, metrics used for outlier rejection can be either computed as a pairwise comparison between all slices (by default) or only on a window of neighboring slices. With all the different variations, we obtain a total of 166 different IQMs.

In addition to the previously described IQMs, we also include a Boolean variable that assesses whether a given IQM computation failed. If this occurs, the IQM will have a zero value and the corresponding Boolean variable will be set to true. This allows to keep all IQMs values to a real number. With the variants and the missing value flag, we reach a total of 332 IQMs. A more thorough description of each IQM used in FetMRQC is available in Table~\ref{tab:metrics} in the supplementary material, along with a cross-correlation matrix on the entire training dataset of the 100 IQMs most frequently used.

\subsubsection{QA/QC prediction}
Given the extracted IQMs, a prediction model is then trained to predict the discrete ratings (QA; regression) or predict whether an image should be excluded (QC; classification), using various machine learning models from the Scikit Learn library~\citep{scikit-learn} and from the XGBoost python package.  For the QA task, we consider linear regression, support vector machine \review{(\texttt{SVR} class using an RBF kernel with a scaled kernel coefficient, regularization parameter C=1.0)}, random forests \review{(\texttt{RandomForestRegressor} class with 100 estimators, fitted using the Gini coefficient)}, and XGBoost’s regression model~\citep{chen2016xgboost} \review{(\texttt{XGBRegressor} class using 100 estimators)}. For the QC task, we consider logistic regression, support vector classifier \review{(\texttt{SVC} class using an RBF kernel with a scaled kernel coefficient, regularization parameter C=1.0)}, random forest \review{(\texttt{RandomForestClassifier} class with 100 estimators, fitted using the Gini coefficient)}, and XGBoost’s classification model~\citep{chen2016xgboost} \review{(\texttt{XGBClassifier} function using 100 estimators). 

Early experiments included also a multi-layer perceptron (\texttt{MLPRegressor} and \texttt{MLPClassifier} classes with multiple hidden layers with up to 1000 neurons per layer), but these models were not found to bring any added value compared to the non deep-learning based approaches, while very largely increasing the training time. They were not used in the following analyses. Note that this behavior is common in tabular data, where deep learning models are not necessarily performing best~\citep{grinsztajn2022tree}.}

 We performed model selection by ablating over the previously mentioned feature normalization and feature selection options, as well as various models.

\paragraph{Pre-processing}
\review{The QA/QC prediction started from the unprocessed clinical acquisitions, converted from the DICOM to the Nifti format. The same pre-processing steps were applied to the data from all the sites considered.}

\textit{IQM normalization.} Domain shifts, also known as batch effects~\citep{leek2010tackling, esteban2017mriqc}, can induce substantial biases in IQM computations. One approach to mitigate them is using group scaling~\citep{esteban2017mriqc}. This is why we experiment with various normalization techniques: standardization, robust (median-based) and quantile scaling, group-wise standardization, group-wise robust/quantile scaling (scaling by subject/scanner/site) and ComBat~\citep{johnson2007adjusting}. In addition to mitigating batch effects, feature standardization is important for models such as logistic or linear regression, but this is not the case for tree-based models.

\textit{Feature selection and dimensionality reduction.} Correlated and irrelevant features can also be an obstacle for machine learning models. We experiment with dropping IQMs that are highly correlated with each other(with thresholds of 0.8 and 0.9), to remove constant features, and experiment with removing features that do not contribute more than noise using the Winnow algorithm~\citep{littlestone1988learning} with extremely randomized trees~\citep{esteban2017mriqc}. Finally, we also explore using principal component analysis to construct orthogonal features. 

\paragraph{Model selection} In our initial experiments, we used nested cross-validation to automatically perform model selection and evaluation without introducing optimistic biases~\citep{varoquaux2017assessing}. We performed model selection by ablating over the previously mentioned feature normalization and feature selection options, as well as the different models. However, in the large majority of these experiments, the best-performing configuration used no standardization, no feature selection, and random forests for both classification and regression. Based on these ablations (available in the Supplementary Material \ref{sec:ablations}), we decided to only use a random forest without standardization or feature selection. As no model selection needs to be carried out, nested cross-validation is not required and will not be used in the rest of the paper.

\subsection{Experimental setting}
 We divide our dataset in two: 1246 stacks were used for training and validation of the models based on cross-validation experiments and 398 were used for assessing the generalization to unseen data, from La Timone and two randomly selected scanners. Data from La Timone were included in the study specifically to serve as external testing from an unseen site. Three increasingly challenging evaluation settings are considered: (i) Subject-wise 10-fold cross validation (CV) on the \textit{training} stacks, which quantifies the expected performance of the method on new subjects acquired on already seen scanners; (ii) Leave-one-Scanner-out (LoSo) CV on the \textit{training} stacks, where each fold leaves out all data from a single scanner for evaluation. This evaluates the expected performance of the method on different scanners; (iii) Pure testing on unseen scanners and an unseen site. This is the closest to a real-world deployment setting, as the pure testing data were not seen during the processes of design and training of the models. 

\paragraph{Baselines} For classification, we consider the following baselines. We first include NiftyMIC-QC~\citep{ebner_automated_2020}, which computes the volume of the brain for each stack and, for each subject, excludes the stacks with a volume below 70\% of the median volume. We also include the deep learning methods of \citet{legorreta2020DL} (\texttt{dl\_stack}) and \citet{xu2020semi} (\texttt{dl\_slice}). These IQMs are computed for each individual subject, we then standardize them and train a logistic regression model to adjust their prediction to the statistics of our dataset. This step adjusts the threshold for prediction and can only be beneficial to the prediction accuracy of these baselines.


For regression, as there is no baseline available to our knowledge, we consider a simple model predicting only subject-wise class statistics for regression, predicting the average rated quality of each subject as quality assessment (e.g. for a subject with three stacks rated as 3.5, 2, 3 respectively, the model assigns the value 2.83 to all stacks). This oracle is based on the assumption that the subject-wise averaged rating can be predictive of the quality rating, which is the case in our data, as the Pearson correlation of the two is $R=0.59$. This method serves as a coarse point of comparison for the QA performance of FetMRQC. 

In addition, for both QC and QA, we assessed the added value of our proposed IQMs as follows. First, we constructed a \textit{Base} version of FetMRQC using the six state-of-the-art IQMs proposed in the context of fetal brain QA/QC. Then, we considered two variants of our model: FetMRQC used all estimated 332 IQMs and FetMRQC-20 used only 20 IQMs (selected based on their measured feature importance on the training data). Note that as this selection was based on the results in evaluation settings (i) and (ii), the performance of the model was likely be inflated due to double dipping~\citep{kriegeskorte2009circular}. It remains nonetheless informative on the expected performance of FetMRQC when only relying on a restricted set of IQMs. FetMRQC-20 is further discussed in our last experiment. All details regarding the baselines is provided in Table~\ref{tab:methods}.

\begin{table}[!t]
    \caption{Summary of the methods compared in the paper.}
    \label{tab:methods}
    \centering
    \resizebox{\linewidth}{!}{
    \begin{tabular}{m{2.cm}P{9cm}}
        \toprule
         \texttt{\textbf{dl\_slice}}&  Slice-wise deep learning (DL) quality control of \citet{xu2020semi}, aggregated into a single score. The decision threshold is learned by logistic regression.\\
         \texttt{\textbf{dl\_stack}}& Stack-wise DL QC of \citet{legorreta2020DL}. The decision threshold is learned by logistic regression.  \\
         \texttt{\textbf{NiftyMIC-QC}}& Subject-wise QC, excluding stacks with brain volume below 70\% of the median brain volume calculated for the subject.\\
         \texttt{\textbf{Base}}&  Base version of FetMRQC using 6 IQMs: rank error~\citep{kainz_fast_2015}, mask centroid~\citep{dedumast2020translating}, mask volume~\citep{ebner_automated_2020}, normalized cross-correlation, mutual information~\citep{kuklisova-murgasova_reconstruction_2012,ebner_automated_2020}, \texttt{dl\_stack}~\citep{legorreta2020DL}, \texttt{dl\_slice}~\citep{xu2020semi}\\
         \texttt{\textbf{FetMRQC}}& Full version of FetMRQC, using 332 IQMs \\
         \texttt{\textbf{FetMRQC-20}}&  Use the 20 best IQMs of FetMRQC -- \texttt{rank\_error}, \texttt{closing\_mask\_full}, \texttt{mask\_volume}, \texttt{filter\_mask\_Laplace}, \texttt{filter\_mask\_sobel\_full}, \texttt{nRMSE\_window}, \texttt{filter\_mask\_Laplace\_full}, \texttt{filter\_mask\_Laplace}, \texttt{closing\_mask}, \texttt{rank\_error\_center}, \texttt{seg\_sstats\_BG\_N}, \texttt{centroid}, \texttt{rank\_error\_center\_relative}, \texttt{seg\_sstats\_CSF\_N}, \texttt{seg\_sstats\_GM\_N},  
         \texttt{im\_size\_z},  \texttt{NCC\_intersection}, \texttt{NCC\_window}, \texttt{PSNR\_window}, \texttt{seg\_SNR\_WM}, \texttt{seg\_volume\_GM}.\\
         \midrule
         \texttt{\textbf{Sub.-wise oracle}} & For each subject, compute the average stack quality and return this value for all the stacks of the subject.\\
         \bottomrule
    \end{tabular}}
    
\end{table}
\noindent\textbf{Evaluation metrics.}
Our classification results use a weighted F1-score, to handle imbalanced classes, and the area under the receiver operating characteristic curve (ROC AUC), as well as precision and recall. Our regression results are evaluated using Pearson’s $R^2$ score, Spearman rank correlation, and mean absolute error (MAE). 

\noindent\textbf{Implementation.} The experiments were implemented with Python 3.9.15 and Scikit-learn 1.1.3~\citep{scikit-learn}. All code is available on Github\footnote{\url{https://github.com/Medical-Image-Analysis-Laboratory/fetmrqc}} and a Docker version\footnote{\url{https://hub.docker.com/u/thsanchez}} is also provided.

\section{Results}
\begin{figure*}[!t]
    \centering
    \includegraphics[width=.9\linewidth]{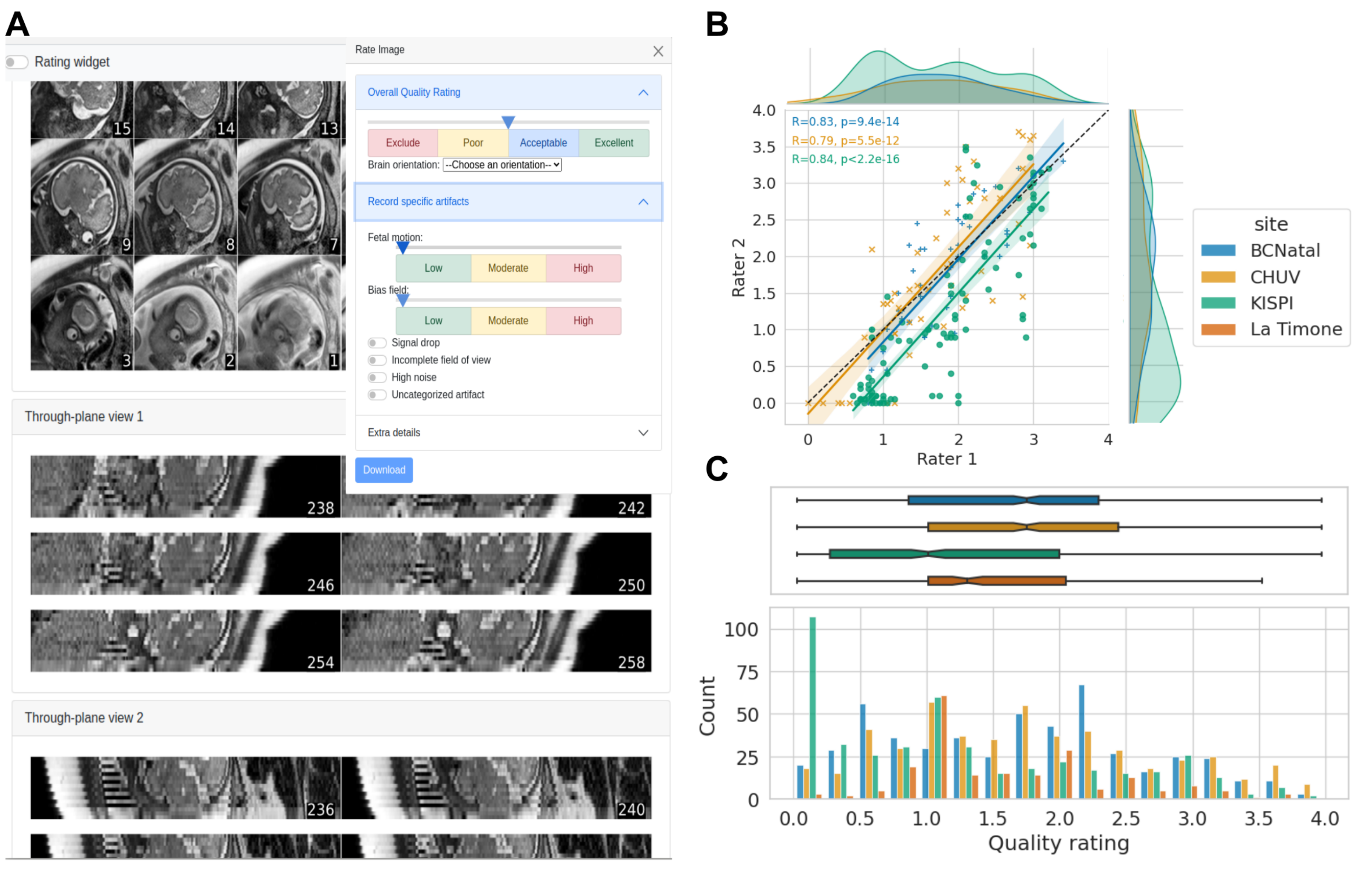}
  \caption{\textbf{A look into the dataset.} \textbf{A} -- Illustration of the quality rating interface developed in this work. \textbf{B} -- Inter-rater agreement on the 211 stacks annotated by both raters. The global R value is 0.75. Note that stacks from La Timone were only annotated by Rater 2. \textbf{C} -- Distribution of the quality ratings across the different sites considered, on all data. The median values are respectively $1.75$ $[0.84,2.4]$ for BCN, $1.75$ $[1, 2.45]$ for CHUV and $1$ $[0.1, 2.05]$ for KISPI.}
      \label{fig:data}
\end{figure*}

\subsection{Stack screening optimization with visual reports}
Using FetMRQC's visual reports interface, Rater 1 annotated 657 stacks, and rater 2 annotated 1203 stacks. 211 of these stacks selected randomly across the training dataset were annotated by both raters to assess inter-rater reliability. Rater~1, YG, is a maternal-fetal physician with 5 years of experience, and Rater~2, MBC, is an engineer with 20 years of experience. The total rating time was 6h 40min for Rater 1 (median of 36s per volume), and 14h20 for Rater 2 (median of 42s per volume).
A high inter-rater agreement was achieved in the manual quality annotations, with Pearson's correlation value of 0.75 overall ($\text{R}^\text{2}$=0.56; Figure~\ref{fig:data}). The inter-rater agreement is consistently high within each site (\ref{fig:data}B). On CHUV data, 127 stacks were manually rated below the exclusion threshold (Quality < 1), and 371 were rated between poor and excellent. On BCNatal data, 155 stacks were excluded, and 353 rated above the threshold. On KISPI, 218 stacks were rated below 1, and 223 above. On La Timone, 42 stacks were rated below 1, and 161 stacks above. The average ratio of excluded stacks is $2.04$. Regarding inclusion and exclusion of stacks (stacks with quality above 1 are included, other are rejected), the inter-rater agreement yielded a Cohen's coefficient of $\kappa=0.58$ (moderate agreement according to the interpretation of \citet{landis1977application}). 

While the raters were trained to rate the overall quality of the images, they also were instructed, but not trained, to rate specific artifacts. They were asked to rate the degree of fetal motion (visible as discontinuities through-plane and signal drops in-plane) and bias field, visible as a low-frequency varying field. However, as their main goal was to give a global rating, the raters often skipped the assessment of the artifacts when the image was either clearly good or clearly bad, leading to inconsistent ratings. for motion rating, their Pearson's correlation drops to $\text{R}^\text{2}$=0.15, and for bias rating, $\text{R}^\text{2}$=0.02. We believe that such a low reliability could be avoided by designing the rating differently, and asking the raters to assess artifacts before giving a global score. In the sequel, we will only use the overall quality rating of the images.

\begin{table*}[!t]
    \centering
    \vspace{0pt}\captionof{table}{\textbf{Quality control and assessment results.} QC (classification, left) and QA (regression, right) results were averaged over five repetitions of the experiment. Results are the median cross-validation performance. The number in parentheses is the average worst-performing cross-validation fold. Three evaluation settings were considered: 10-fold subject-wise cross-validation (CV), LoSo CV and pure testing. Pure testing evaluation was grouped by scanners in the testing set.}\label{tab:vanilla_perf}
    \resizebox{.6\linewidth}{!}{
    \begin{tabular}{lcccc}
    \toprule
    
    \multicolumn{5}{c}{\textsc{Quality control (classification)}}\\
     & Weighted F1 ($\uparrow$)  & ROC AUC ($\uparrow$) & Precision ($\uparrow$) & Recall ($\uparrow$)\\
    \midrule
    \multicolumn{5}{c}{10-fold subject-wise cross-validation}\\
    \midrule 
    
    \texttt{dl\_slice}~\citep{xu2020semi}       & $0.64$ ($0.65$)& $0.72$ ($0.61$)& $0.71$ ($0.73$)& $0.98$ ($0.86$)\\
    \texttt{dl\_stack}~\citep{legorreta2020DL}  &$0.71$ ($0.72$)& $0.77$ ($0.73$)& $0.78$ ($0.80$)& $0.85$ ($0.81$) \\
    NiftyMIC-QC~\citep{ebner_automated_2020}         & 0.76 (0.75) & -- & 0.76 (0.77) & 0.96 (0.96)\\
    Base                                            & $0.82$ ($0.78$)& $0.88$ ($0.79$)& $0.85$ ($0.83$)& $0.92$ ($0.84$) \\
    FetMRQC                                         & $0.86$ ($0.79$)& $0.91$ ($0.87$)& $0.86$ ($0.85$)& $0.94$ ($0.86$)\\
    FetMRQC-20                                  & $0.86$ ($0.77$)& $0.92$ ($0.87$)& $0.86$ ($0.85$)& $0.93$ ($0.81$) \\
    \midrule 
    \multicolumn{5}{c}{Leave-one-Scanner-out cross-validation}\\
    \midrule 
    \texttt{dl\_slice}~\citep{xu2020semi} & $0.61$ $(0.47)$& $0.75$ $(0.60)$& $0.70$ $(0.62)$& $0.96$ $(0.93)$ \\
    \texttt{dl\_stack}~\citep{legorreta2020DL} & $0.64$ $(0.53)$ & $0.75$ $(0.62)$ & $0.69$ $(0.47)$& $0.90$ $(0.87)$ \\
    NiftyMIC-QC~\citep{ebner_automated_2020}   & 0.75 (0.66)  & -- & 0.76 (0.71) & 0.95 (0.86)\\
    Base & $0.78$ $(0.63)$& $0.80$ $(0.76)$& $0.80$ $(0.69)$&$0.84$ $(0.67)$ \\
    FetMRQC & $0.80$ $(0.64)$& $0.89$ $(0.74)$& $0.85$ $(0.71)$& $0.86$ $(0.73)$ \\
    FetMRQC-20 & $0.82$ $(0.72)$& $0.90$ $(0.83)$& $0.85$ $(0.76)$& $0.88$ $(0.83)$ \\
    \midrule
    \midrule
    \multicolumn{5}{c}{Pure testing (KISPI + CHUV + La Timone -- by scanner)}\\
    \midrule 
    \texttt{dl\_slice}~\citep{xu2020semi} & 0.73 ($0.76$) & 0.79  ($0.79$) &0.77 (0.77) &0.97 (0.92)  \\
    \texttt{dl\_stack}\review{~\citep{legorreta2020DL}} & 0.62 ($0.60$) & 0.72  ($0.51$) & 0.68 (0.67) & 0.97 (0.86) \\
    NiftyMIC-QC~\citep{ebner_automated_2020} &0.74 (0.52) & -- &0.70 (0.65)  &0.98 (1.00) \\
    Base &$0.77$ ($0.54$) & $0.77$  ($0.62$) & $0.80$ (0.65) & $0.97$ (1.00)\\
    FetMRQC & $0.82$ ($0.67$) & $0.77$  ($0.76$) & $0.83$ (0.70)& $0.91$ (0.91)\\
    FetMRQC-20 & $0.79$ ($0.56$) & $0.74$  ($0.64$) & $0.78$ (0.65)& $0.93$ (0.94)\\
    \bottomrule
    \end{tabular}}
    \resizebox{.38\linewidth}{!}{
    \begin{tabular}{lccc}
    \toprule
    
    \multicolumn{4}{c}{\textsc{Quality assessment (regression)}}\\
     & $R^2$ ($\uparrow$)  & Spearman ($\uparrow$) & MAE ($\downarrow$) \\
    \midrule
    \multicolumn{4}{c}{10-fold subject-wise cross-validation}\\
    \midrule 

    Subject-wise oracle & 0.33 (0.39) & 0.53 (0.68) & 0.65 (0.61) \\
    Base & $0.40$ (0.38) & $0.69$ (0.68) & $0.59$ (0.61) \\
    FetMRQC & $0.60$ (0.49) & $0.80$ (0.75) & $0.50$ (0.56)\\
    FetMRQC-20 & $0.60$ (0.53) & $0.79$ (0.78) & $0.50$ (0.53) \\
    \midrule 
    \multicolumn{4}{c}{Leave-one-Scanner-out cross-validation}\\ 
    \midrule
    Subject-wise oracle & 0.29 (0.40) & 0.48 (0.58) & 0.64 (0.64) \\
    Base & $0.29$ (0.25) & $0.59$ (0.48) & $0.64$ (0.66) \\
    FetMRQC & $0.45$ (0.39) & $0.74$ (0.72) & $0.56$ (0.60)\\
    FetMRQC-20 & $0.52$ (0.36) & $0.77$ (0.71) & $0.55$ (0.62) \\
    \midrule
    \midrule
    \multicolumn{4}{c}{Pure testing (KISPI + CHUV + La Timone -- by scanner)}\\ 
    \midrule 
    Subject-wise oracle & 0.41 (~0.41) & 0.60 (0.60) & 0.45 (0.45) \\
    Base &  $0.26$ (~0.36)&  $0.45$ (0.47)& $0.65$ (0.37)\\
    FetMRQC & $0.35$ (-0.74)& $0.59$ (0.39)&  $0.51$ (0.65)\\
    FetMRQC-20 & $0.30$ (-0.94)& $0.54$ (0.31)& $0.53$ (0.68)\\
    \bottomrule
    \end{tabular}}
    \end{table*}

\subsection{Performance and robustness of FetMRQC}
\begin{figure}[!t]
    \centering
    \includegraphics[width=\linewidth, trim=2 2 2 2,clip]{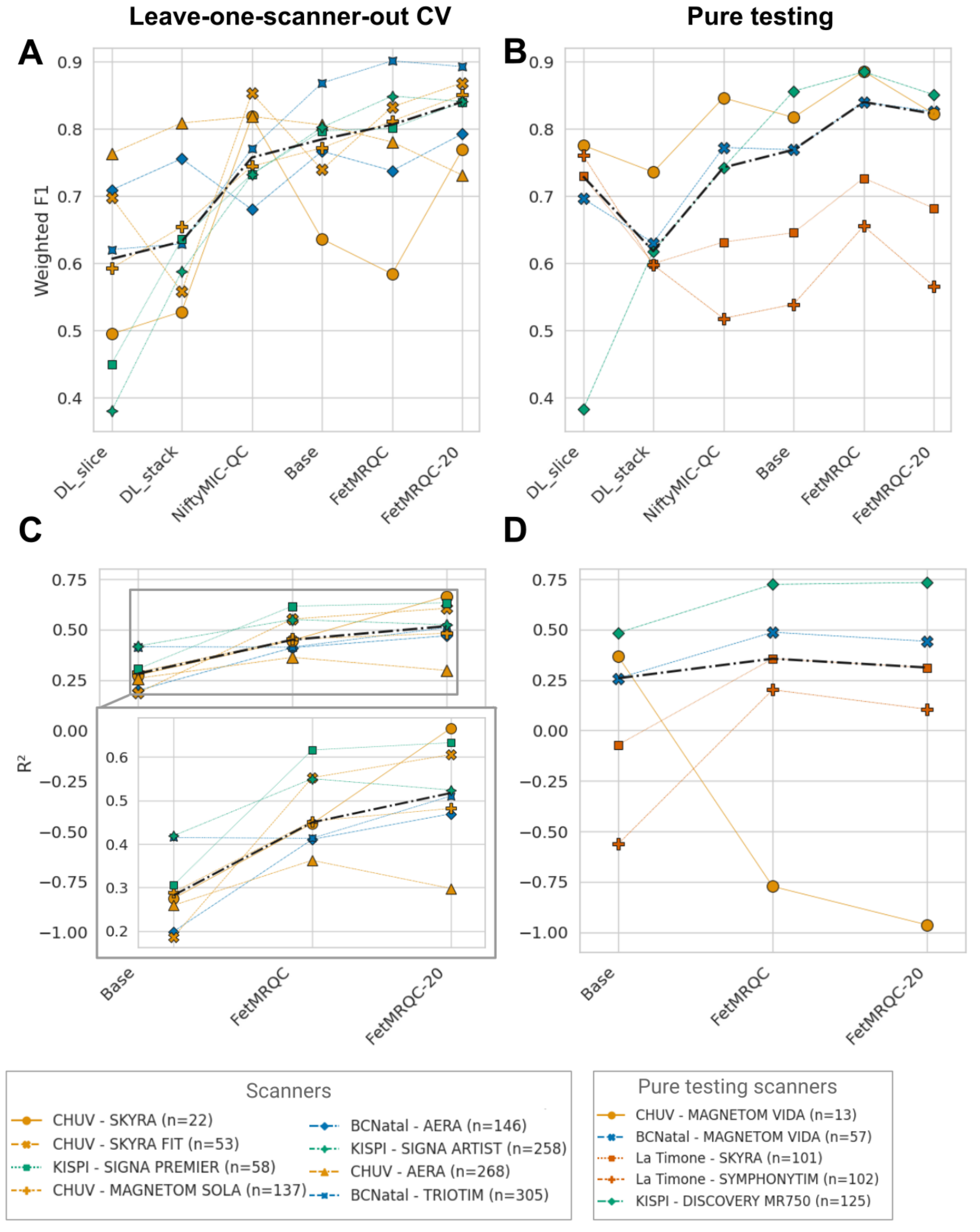}
    \caption{\textbf{Scanner-wise results for QA/QC.} \textbf{A} -- Weighted F1 score for the QC task for each scanner used in LoSo cross-validation (sorted from the one with the least subjects to the most subjects). \textbf{B} -- Weighted F1 score for the QC task for each scanner used in the pure testing set. \textbf{C} -- $\text{R}^\text{2}$ for the QA task for each scanner used in LoSo cross-validation. \textbf{D} -- $\text{R}^\text{2}$ for the QA task for each scanner used in the pure testing set. Distribution of scores is aggregated by scanner, and the median performance for each method is shown as the black dashed line. The red line in the prediction task at $0$ shows the baselines for a constant predictor. These results detail the ones presented in Table \ref{tab:vanilla_perf}.}\label{fig:boxplot_scanner}
\end{figure}

Based on the ratings from FetMRQC, we considered two tasks: a quality control (QC) task, where we aimed at predicting whether a scan should be excluded (rating below 1), and a quality assessment (QA) task, where we predicted the interval rating (between 0 and 4). Results from the experiment are summarized in Table~\ref{tab:vanilla_perf}.
A more detailed outlook at the variations in performance across scanners in the LoSo cross-validation and pure testing performance is available in Figure~\ref{fig:boxplot_scanner}. As expected, the three increasingly challenging evaluation settings (10-fold CV, LoSo CV, pure testing) led to a decrease of performance. This decrease is less notable for QC than QA. 

\noindent\textbf{Quality control.} Overall, FetMRQC and FetMRQC-20  consistently performed best with a performance (weighted F1) of 0.86, 0.80 and 0.82 in median for the cross-validation, leave-one-out scanner and pure testing scenarios respectively. This performance is consistent across the evaluation metrics considered (\ref{tab:vanilla_perf}). Precision is of great interest in our case, as including bad quality in further analysis can be greatly detrimental to further processing. FetMRQC shows a consistently high precision in all settings considered, with median performance of 0.86, 0.85 and 0.83 in CV, LoSo CV and pure testing respectively. 

Focusing on the scanner-wise breakdown of performance (Figure \ref{fig:boxplot_scanner}A and B), FetMRQC and FetMRQC-20's performance is very consistent across almost all scanners considered, and does not change on new scanners from sites used in training (Siemens' MAGNETOM Vida at CHUV and BCNatal - GE's Discovery MR750 at Kispi). On the other hand, DL-based methods~\citep{legorreta2020DL,xu2020semi}, trained on homogeneous data from a single site, fail to perform and exhibit very large variations in performance across sites, making them generally unreliable. We note also that a few scanners were consistently challenging for the models. On panel A, we see that all methods except NiftyMIC-QC and FetMRQC-20 struggled on the CHUV - Skyra scanner. On panel B, we see that FetMRQC managed to generalize well to unseen scanners from known sites (BCN, KISPI and CHUV). However, all models, except \texttt{dl\_slice}, poorly generalized to data from La Timone.

\noindent\textbf{Quality assessment.} In the case of quality assessment, we observed that FetMRQC's new IQMs were instrumental in achieving a performance above the subject-wise oracle. On Table~\ref{tab:vanilla_perf}B, we see that while the IQMs used in the base model ($\text{R}^2$=0.49) were sufficient to outperform the subject-wise oracle ($\text{R}^2$=0.33) in the subject-wise CV, using FetMRQC with either all IQMs ($\text{R}^2$=0.44) or the selected 20 ($\text{R}^2$=0.49) was necessary to achieve a performance over the subject-wise oracle ($\text{R}^2$=0.29) in the LoSo setting. This was nonetheless not sufficient to achieve a satisfying performance in the pure testing setting, where FetMRQC's prediction, despite outperforming consistently over the base model, do not outperform the subject-wise oracle. It also fails on one scanner (CHUV - MAGNETOM Vida scanner, Figure \ref{fig:boxplot_scanner}D), but we hypothesize that such drop is likely due to the small amount of data available from this scanner.

\begin{figure}[!t]
    \centering
    \vspace{0pt}\includegraphics[width=\linewidth]{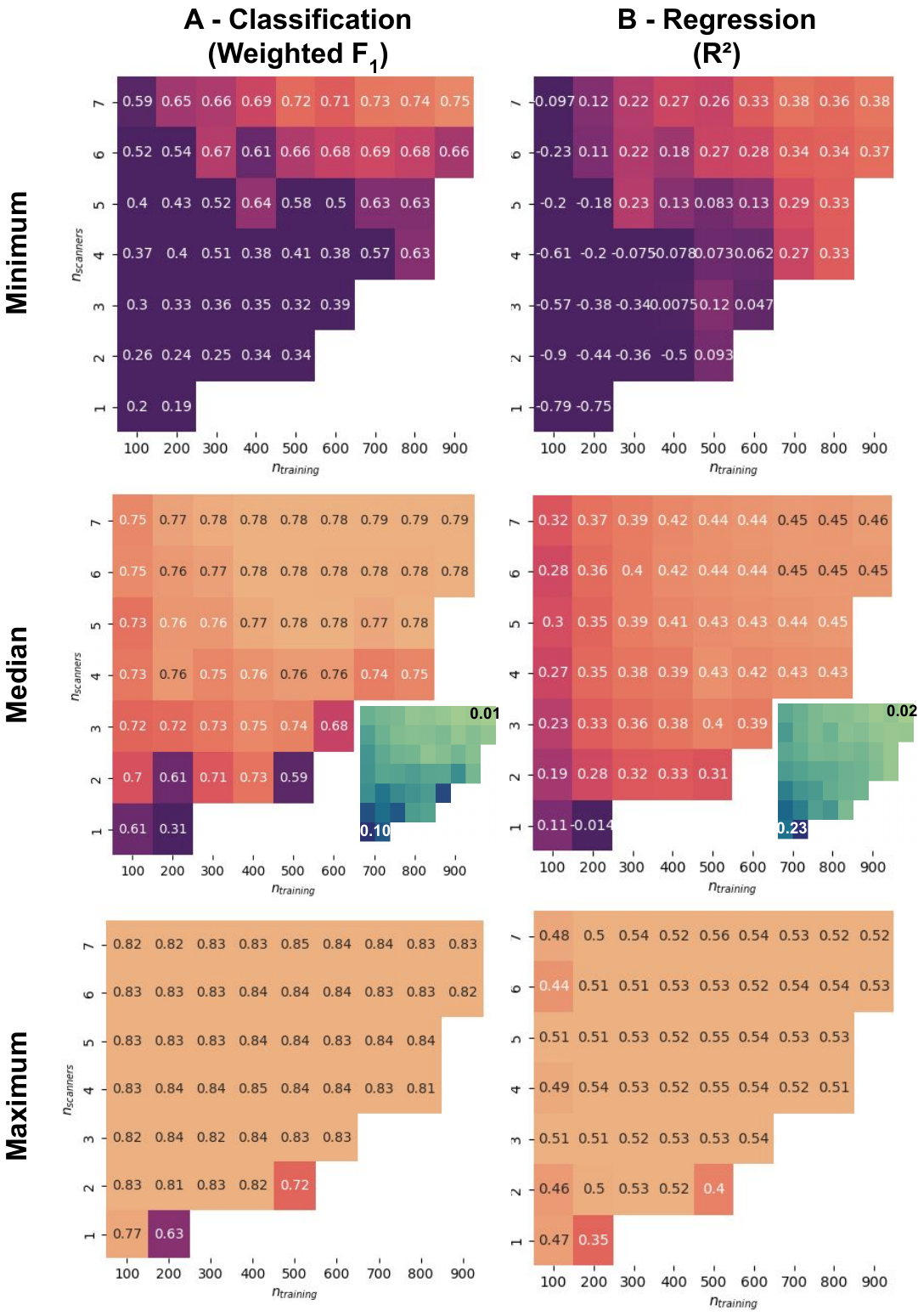}
    
  \caption{\textbf{Performance as a function of the number of scanners and training points.} This is obtained by performing leave-one-scanner-out cross-validation 20 times, using different random subsets of data. \textbf{(Top row.)} Minimum (worst-case) performance across folds \textbf{(Middle row.)} Median performance across folds. The smaller plots show the corresponding median average deviation. \textbf{(Bottom row.)} Maximum (best-case) performance across folds.}
      \label{fig:train_size}
\end{figure}

\subsection{Generalization as a function of scanner diversity and number of training examples}
Data annotation is known to be a time-consuming process that requires highly specialized raters~\citep{radsch2023labelling}. Given a limited budget (in time and expertise), the question of which data to annotate then raises naturally. In this experiment, we investigated how the number of scanners $n_\text{scanner}$ and the number of data  $n_\text{training}$ available during training impacted the generalization performance of FetMRQC in the context of LoSo CV. We had in total 8 different scanners and 1251 data points. For a given configuration ($n_\text{scanner}$, $n_\text{training}$), we performed a LoSo CV where the data used in training were subsampled: between 1 and 7 scanners were sampled randomly from the available data and between 100 and 900 data points were then randomly sampled from the available scanners. For each ($n_\text{scanner}$, $n_\text{training}$), the experiment was repeated 20 times. 

Figure~\ref{fig:train_size} contains the results of the experiment, showing the minimum, maximum and median performance with the  deviation from the median, across 20 repetitions. In each case entry, the reported measure was computed as the average across the 20 repetitions. Looking at the median performance, it is clear that increasing the size of the training set (x axis) or the number of scanners (y axis) both improve the generalization. Starting with best-case generalization (maximum performance, lower row in figure \ref{fig:train_size}), we see that in every case, there is a subset of data that enables reaching the best performance with only 100 data points. 
While this is not surprising, this is also difficult to exploit: one cannot readily find ahead of time a subset of data that will generalize well to the testing data. 
The worst-case generalization is more interesting: using 100 training data points from seven scanners reaches a similar performance as using 700 data points from four scanners in the case of classification. In the case of regression however, we see that both the number of training samples and scanners is important: the worst-case generalization with 100 training data and 7 scanners is close to zero, and the performance steadily increases with more data.

Overall, using multiple scanners is key to achieving the highest performance regimes, but using more data is also greatly valuable. However, if constrained to a limited annotation budget, we anticipate that annotating more diverse data from various scanners will be more helpful for generalization than gathering a large corpus from a single scanner.

In addition, we also observe, on the median performance, that the classification task is generally more straightforward than the regression task: fewer data allow to reach the highest performance, while performance keeps increasing for regression when adding more sites and more data. Thus, we hypothesize that regression performance would further be increased by increasing the size of training data. In contrast, the median classification performance might stagnate, although its worth case performance might still improve, thus making the model more robust to new scanners by further enhancing the training dataset.

\subsection{Model performance on a restricted set of IQMs}
FetMRQC relies 332 different IQMs that are not fully independent from each other, as shown in Figure \ref{fig:crosscorr}. In this final experiment, we explore the IQMs that are most important for FetMRQC QA and QC models. 

\begin{figure*}[!t]
    \centering
    \includegraphics[width=0.7\linewidth]{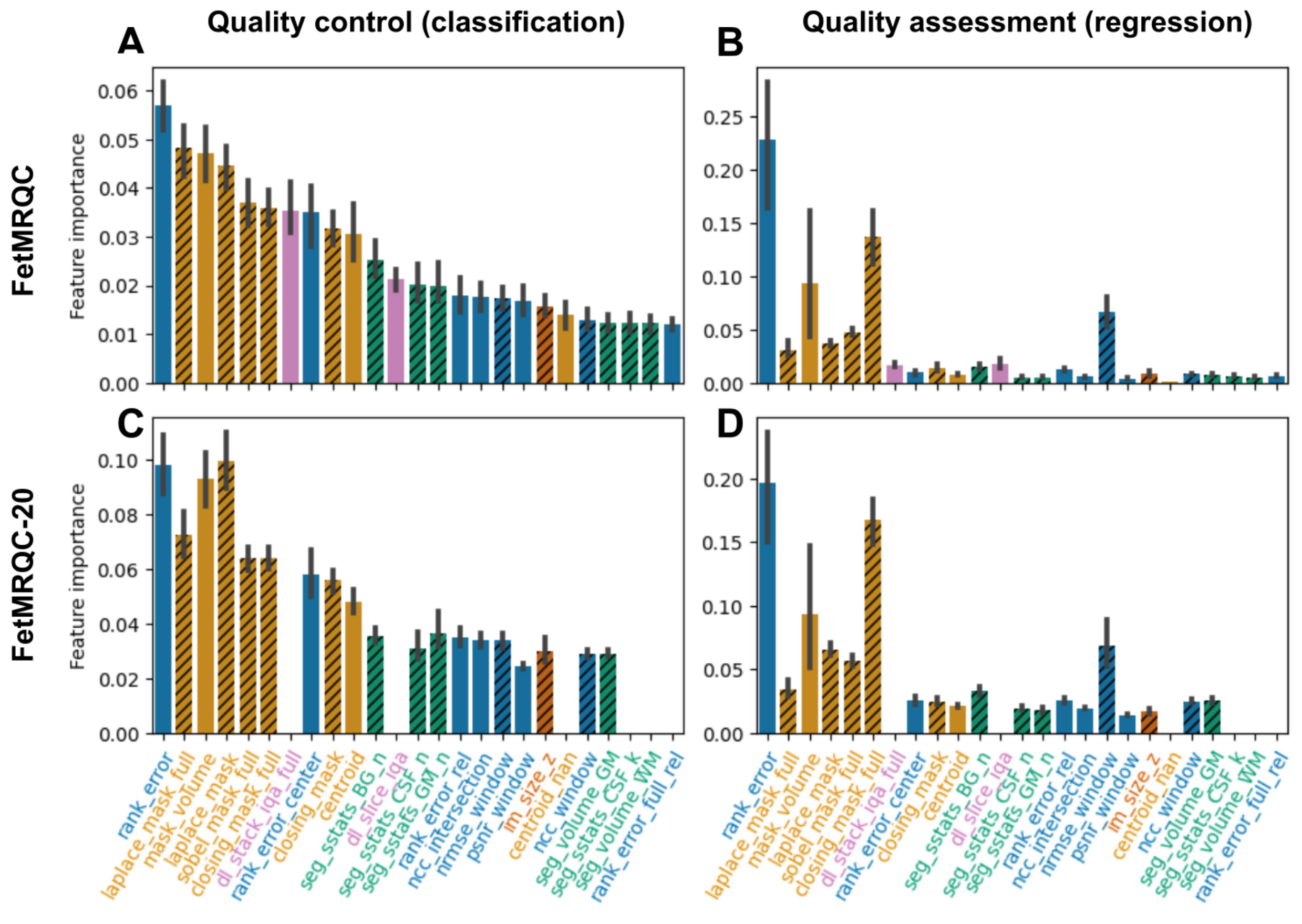}
    
  \caption{\textbf{Most important IQMs for QA/QC.} Feature importance for quality control (classification) on the left, and for quality assessment (regression) on the right. The top row shows the top-25 IQMs from FetMRQC and the bottom row shows the 20 selected IQMs that form FetMRQC-20. Blue IQMs are intensity-based, orange are mask- (or shape) based, green are segmentation based, pink are deep-learning based and brown are metadata based. Hatched features denote the new ones proposed in this work. The error bars are the standard deviation over the different cross-validation folds, performed over different scanners. Note that the scales are very different between the plots: the highest feature importance for classification is around 0.055, whereas it is around 0.23 for regression.}
      \label{fig:iqms}
\end{figure*}

We computed the feature importance of the random forest model used in each fold of the LoSo CV and average them across folds. We grouped together the IQMs with a correlation coefficient above 0.95 (as shown in Figure \ref{fig:crosscorr}) to prevent several IQMs contributing very similar information but selected by different models in the LoSo CV for QA and QC. We then randomly selected a single IQM from each correlated group, and arrived at the ranking shown in the top row of Figure \ref{fig:iqms}. First, we see that in the QC task (A), IQMs are generally spread out (the top four IQMs sum up to 0.20). In the regression task (B) however, a few IQMs capture a large part of the feature importance (the top four IQMs sum up to 0.53). Nonetheless, three IQMs are consistently among the top predictors: rank-based error~\citep{kainz_fast_2015}, the volume of the brain mask and the morphological closing of the brain mask. The first estimates the consistency of the intensities across slices by computing how well a low-rank approximation can represent the volume, the second estimates the volume of the brain and the third estimates the degree of motion across stacks by computing a morphological closing of the brain mask in the through-plane direction and then subtracting the original brain mask. The first two IQMs are the ones that have been used in NiftyMIC-QC~\citep{ebner_automated_2020} and complement each other well.
Secondly, we see that although the ranking of the most important IQMs can vary,  overall 19 out of the 25 IQMs of Figure~\ref{fig:iqms}A and B appear in common in both tasks as the most important IQMs. Thirdly, let us note that the best IQMs cover different representative families of features: intensity-based, mask (or shape)-based, and segmentation-based IQMs. Finally, note that features proposed within FetMRQC rank highly in terms of feature importance: 14 out of the 25 IQMs shown in Figure~\ref{fig:iqms}A and B were proposed in this work. 

FetMRQC-20 is built on the feature importance obtained for FetMRQC (Figure~\ref{fig:iqms}A and B). The IQMs were selected by averaging the feature importance from QC and QA, and then by selecting the top-20 features. In order to keep the reduced model as interpretable as possible, we excluded the deep learning (DL)-based IQMs from FetMRQC-20 and replaced them with the two features that came next in line. Results in Table~\ref{tab:vanilla_perf} show that does not yield a decrease in performance. The feature importance using only FetMRQC-20's IQMs is shown on Figure~\ref{fig:iqms}C and D and is generally consistent with FetMRQC's results. As fewer IQMs are available, their relative importance is generally higher, and the same IQMs end up carrying the largest weight in decision.

\section{Discussion}
In this work, we proposed FetMRQC, a novel open-source machine learning framework for the automated quality control and quality assessment of fetal brain MRI. While most existing works focus on a single-center, single-scanner setting~\citep{legorreta2020DL,xu2020semi,gagoski2022automated}, the evaluation in this work was carried out on a large, multi-scanner, multi-centric dataset. These diverse data allowed us to measure the impact of domain shift on generalization, and assess the variability in performance across scanners. Being trained with multi-centric data FetMRQC achieves a reliable performance in quality control over most scanners considered, which is not the case for baseline DL-methods, trained on homogeneous data, which exhibit a very large variability in performance. These observations were made possible by following good practices regarding evaluation and reporting of dataset with domain shifts~\citep{roberts2021common,varoquaux2022machine,zech2018variable}. Indeed, cross-validation at the group level (subject or scanner in our case)~\citep{varoquaux2017assessing}, computing the performance metrics at the group level and reporting the worst-performing site were essential in unfolding the large variability in performance, which is obfuscated when averaging across the entire testing set~\citep{dockes2021preventing,zhou2022domain}. Designing a pure testing~\citep{varoquaux2022machine,kapoor2022leakage} set comporting both unseen scanners and unseen sites allowed to observe another trend: methods performing well in the LoSo CV setting performed well on unseen scanners from known sites, but struggled on the unseen site. Indeed, data from La Timone were very different from the ones acquired at other institutions: only three stacks were acquired per subject due to strong constraints on the duration of the scanning session, and the acquisition was done at a high in-plane resolution, leading to higher level of noise in the images compared to the rest of the data. 

Beyond measuring the impact of domain shifts, several methods to correct and compensate them on tabular data have been proposed, including group-wise normalization of data~\citep{esteban2017mriqc}, or empirical Bayes approaches, like ComBat~\citep{johnson2007adjusting}. As shown in our supplementary experiments, we did not find these approaches to be beneficial in our case, which is most likely due to the quality of data being related to the scanner on which data were acquired: removing the scanner information at the IQM level might not be helpful because it might remove meaningful information~\citep{dockes2021preventing}. This might be mitigated by attempting to directly harmonize the input T2w images~\citep{zhou2022domain,wang2022generalizing} rather than the IQMs, as the IQMs were directly  extracted from images acquired with widely different imaging parameters that could induce some confounding factors in the derived metrics.

\review{A question that can be raised is whether a deep models (like convolutional neural networks (CNN) or transformers~\citep{vaswani2017attention}) could serve as an alternative to FetMRQC. FetMRQC operates in a highly heteorogeneous setting, with relatively few, high dimensional data points when compared to deep learning standards -- where datasets commonly feature more than $10^5-10^6$ data points \citep{deng2009imagenet,varoquaux2022machine}. Using our data, we were unable to train a CNN or a transformer model that would outperform FetMRQC. In addition, the trained models exhibited unstable generalization performances. We hypothesize that the diversity of IQMs of FetMRQC, leveraging image intensity, brain masks and finer segmentations were able to provide a more stable ground for generalization than the one learned by a deep learning model on our data. Our choice of privileging random forests over deep networks in FetMRQC then hinged on practical considerations, rather than the theoretical representation power of deep networks. Nevertheless, deep learning has still been successful for quality control~\citep{legorreta2020DL,xu2020semi,liao2020joint} and it is likely that having more data or leveraging semi-supervised~\citep{xu2020semi} or self-supervised~\citep{liu2021self,he2022masked} learning methods could help build some robust deep models.}


\review{Note however that} FetMRQC suffers from two main limitations.
As any other supervised learning method, the first limitation comes from an often underestimated component of machine learning pipelines, namely the quality of annotations. As QA/QC has an inherently subjective dimension, narrowing the task at hand for rating is key to maximize inter-rater reliability~\citep{esteban2018improving,radsch2023labelling}. The quality rating interface is an essential tool for displaying the raw T2w fetal brain data uniformly, and when providing the raters with a training session, can successfully lead to high inter-rater reliability. However, our fetal motion and bias field rating results suggest that a finer protocol is needed. The protocol should, in particular, encourage raters to proceed in artifact-based quality ratings: first assessing the presence and degree of various artifacts and then deciding on a score to give rather than the opposite. Improving the inter-rater agreement might further improve the quality of FetMRQC, in particular on the quality assessment task, where the subject-wise CV regression performance comes close to the level of agreement between the raters: $\text{R}^\text{2}$=0.58 for the subject-wise CV and the inter-rater agreement has $\text{R}^\text{2}$=0.56. A second limitation comes from the simplicity of the model: while FetMRQC's predictions are easily interpretable and generally depend on a small number of IQMs, its learning capabilities are limited by its shallow nature. A deep learning model trained directly on 3D clinical acquisitions is likely to improve QA/QC predictions, if enough training data is available, as it can make better  use of large amounts of training data.

Beyond addressing these limitations, future work will investigate how preprocessing the raw T2w data might impact FetMRQC's performance. Future work will also include a more thorough evaluation of the impact of FetMRQC on downstream tasks such as super-resolution reconstruction quality. FetMRQC is only a first step towards robust tools for quantitative analysis of fetal neuroimaging. While QA/QC starts at the raw images, it is greatly needed at every stage of the fetal brain MRI pipeline, from acquisition to reconstruction to surface extraction. Such checkpoints, along with community efforts in collecting large, reality-centric datasets are key to developing robust and reliable learning-based approaches for fetal neuroimaging and beyond.


\section*{CRediT authorship contribution statement}

\begin{itemize}[noitemsep]
\item[] \textbf{Conceptualization:} MBC, TS, OE 
\item[] \textbf{Data:} EE, AJ, NG, MK, VD, GA 
\item[] \textbf{Annotations:} YG, MBC
\item[] \textbf{Methodology:} TS, MBC, OE  
\item[] \textbf{Software:} TS
\item[] \textbf{Evaluation:} TS, AP
\item[] \textbf{Supervision:} MBC
\item[] \textbf{Writing—original draft:} TS, MBC, OE, GA
\item[] \textbf{Writing—review \& editing:} All authors
\end{itemize}

\section*{Declaration of competing interest}
The authors declare that they have no known competing financial interests or personal relationships that could have appeared to influence the work reported in this paper.

\section*{Data availability}
Raw fetal brain MRI cannot readily be shared because of patient privacy. Derived measures, such as extracted IQMs are available on Zenodo (under CC BY 4.0 license) at \url{https://zenodo.org/records/10118981} and on GitHub (under an Apache 2.0 license) respectively, for the results to be reproduced.

\section*{Acknowledgments}
TS and MBC acknowledge access to the facilities and expertise of the CIBM Center for Biomedical Imaging, a Swiss research center of excellence founded and supported by CHUV, UNIL, EPFL, UNIGE and HUG. TS acknowledges support from Era-net Neuron MULTIFACT -- Swiss National Science Foundation (SNSF) grant 31NE30\_203977, OE is supported by SNSF \#185872, National Institute of Mental Health (NIMH) RF1MH12186, Chan Zuckerberg Initiative (CZI) EOSS5-0000000266. GA, AP acknowledge support from ERA-net NEURON MULTIFACT -- French National Research Agency, Grant ANR-21-NEU2-0005 and French National Research Agency, SulcalGRIDS Project, Grant ANR-19-CE45-0014. YG acknowledges support from the SICPA foundation and EE from Instituto de Salud Carlos III (ISCIII) grant AC21\_2/00016. \review{AJ is supported by the Prof. Max Cloetta Foundation, EMDO Foundation and Vontobel Foundation.}

\bibliographystyle{abbrvnat}
{\small\bibliography{biblio}}

\newpage
\FloatBarrier
\onecolumn
\section{Supplementary material}
\subsection{IQMs}

\begin{figure}

\centering
    \includegraphics[width=\linewidth]{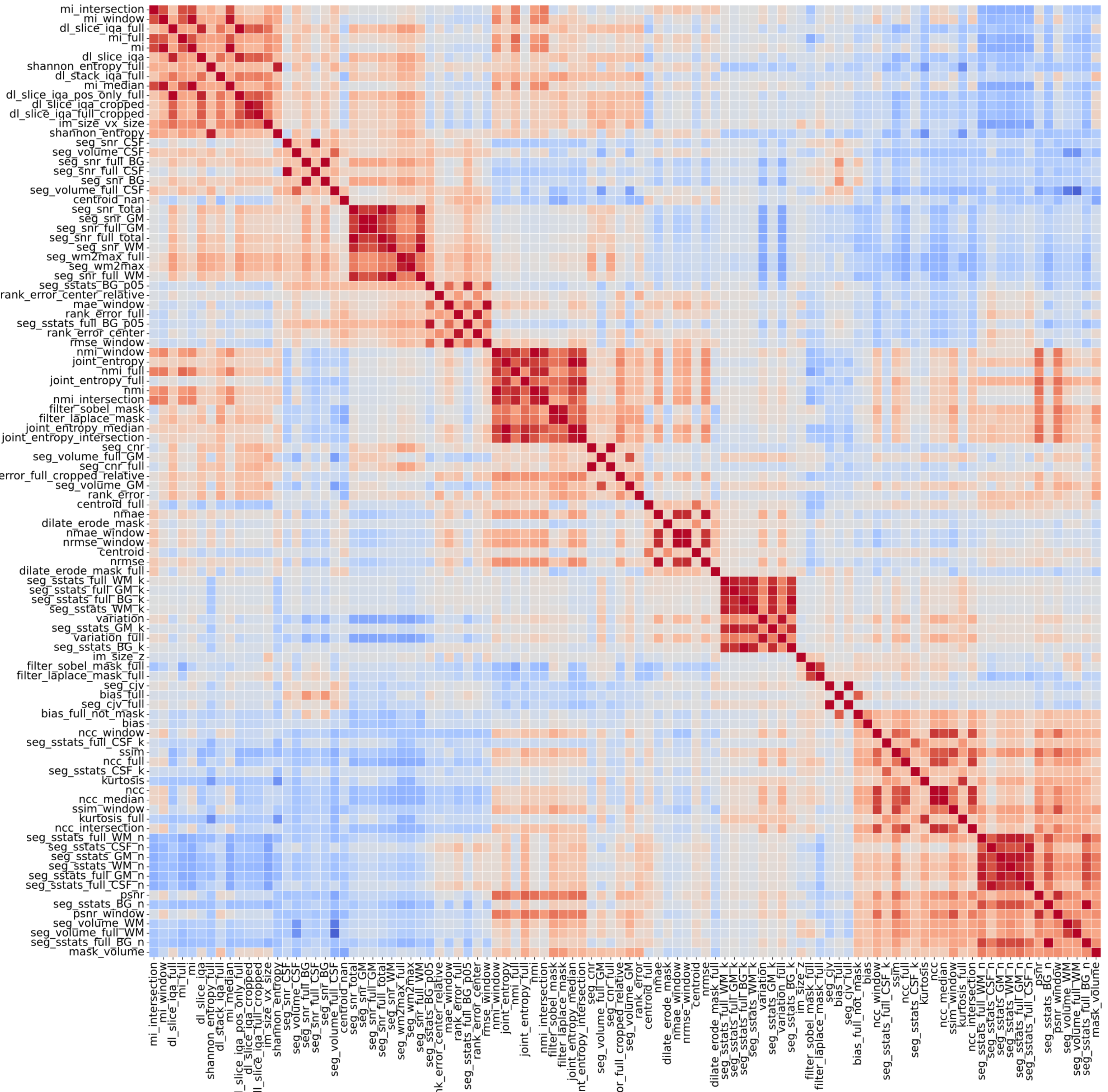}
    \captionof{figure}{Correlation matrix between the top 100 features used in FetMRQC, according to feature importance, evaluated on the entire dataset. Blue refers to negative correlations, and red to positive ones. The scale goes between -1 and 1. The features are clustered by similarity.}
    \label{fig:crosscorr}
\end{figure}

We provide additional details on the image quality metrics used in this work in Table~\ref{tab:metrics}, as well as a cross-correlation matrix of the 100 IQMs most frequently used by FetMRQC (in terms of feature importance) on Figure~\ref{fig:crosscorr}. Table~\ref{tab:metrics} provides a detailed allocation of the 332 IQMs and how they are split between intensity-based, mask-based, segmentation-based, deep learning-based and metadata-based categories.  Figure \ref{fig:crosscorr} shows that although these IQMs tend to cluster different groups, they remain generally independent from each other, and so can serve as complementary information for FetMRQC's prediction model. Highly correlated IQMs tend to be variants of each other: the variant with center-most slices can be very close to the full image IQM (\text{\_full} in the Figure), even if this is not systematically true. Clusters can happen also with IQMs denoting similar quantities: kurtosis in the segmented tissues (\texttt{\_k} in the Figure), or number of voxels in the segmented classes (\texttt{\_n} in the Figure).

\begin{table}[!ht]
    \centering
    \captionof{table}{\textbf{Detailed description of the Image Quality Metrics (IQMs) computed in FetMRQC.} The number in parentheses are the total available variants on each metric (e.g. computation on the masked image, on the central slices, etc.). The number of IQMs sums up to 166 variants. The final number is doubled by incorporating an indicator variable of whether a given entry failed to be computed, resulting in a NaN (not-a-number). }
    \resizebox{.7\linewidth}{!}{
    \begin{tabular}{m{5cm}m{.5cm}m{12cm}}
    \toprule
    \multicolumn{3}{l}{\textsc{Intensity-based metrics}}\\
    \midrule
    \texttt{rank\_error}~\citep{kainz_fast_2015} & (5) & Measure the compressibility of the image using a low-rank approximation\\
    \texttt{slice\_loss} & (32) & Use metrics commonly used for outlier rejection~\citep{kuklisova-murgasova_reconstruction_2012,kainz_fast_2015,ebner_automated_2020} to compute the difference between slices in the volume. We considered (normalized) mean averaged error, (normalized) mutual information, normalized cross correlation, (normalized) root mean squared error, peak signal-to-noise ration, structural similarity and joint entropy.\\
    \texttt{sstats}~\citep{esteban2017mriqc}  & (14)    & Compute the mean, median, standard deviation, percentiles $5\%$ and $95\%$, coefficient of variation and kurtosis on brain ROI.\\
     \texttt{entropy}~\citep{esteban2017mriqc} & (2) & Measure the overall entropy of the image.\\
    \texttt{bias} & (3) & Level of bias estimated using N4 bias field correction~\citep{tustison_n4itk_2010} \\
    \texttt{filter\_image} & (4) & Estimate the sharpness by using Laplace and Sobel filters (commonly used for edge detection)\\
    \midrule
    \multicolumn{3}{l}{\textsc{Mask-based metrics}}\\
    \midrule
    \texttt{mask\_volume} & (1) & Compute the volume of the brain mask.\\
    \texttt{centroid}~\citep{dedumast2020translating} & (2) & Measure the variance in the center of mass of the brain mask across slices.\\
    \texttt{closing\_mask} & (2) & Morphological closing of the brain mask in the through-plane direction, to detect inter-slice motion. Report the average difference with the original mask.\\
    \texttt{filter\_mask} & (4)& Estimate the sharpness of the brain mask using Laplace and Sobel filtering. In an ideal case, the brain mask would be smoothly varying, especially in the through-plane direction.\\
    \bottomrule
    \midrule
    \multicolumn{3}{l}{\textsc{Segmentation-based metrics}~\citep{esteban2017mriqc}}\\
    \midrule
    \texttt{sstats} & (64) & Summary statistics on each region of the segmentation (white matter (WM), gray matter~(GM) and cerebrospinal fluid (CSF). Computing mean, median, 5th and 95th percentile, kurtosis, standard deviation, mean absolute deviation and number of voxels)\\
    \texttt{volume} & (6) & Volume of the three brain regions (WM, GM, CSF; entire brain/central slices)\\
    \texttt{SNR}~\citep{dietrich2007measurement} & (10) & Signal-to-noise computed in each region (background, WM, GM, CSF and globally)\\
    \texttt{CNR}~\citep{magnotta2006measurement} & (2) & Contrast-to-noise-ratio, to estimate the separation between GM and WM.\\
    \texttt{CJV}~\citep{ganzetti2016intensity} & (2) & Coefficient of joint variation of GM and WM.\\
    \texttt{WM2Max} & (2) & White-matter to maximum intensity ratio.\\
    \midrule
    \multicolumn{3}{l}{\textsc{Deep learning-based metrics}}\\
    \midrule
    \texttt{dl\_slice}~\citep{xu2020semi} & (5) & Slice-wise deep learning-based quality assessment. Several variants are considered: full image/central-slices, uncropped/cropped image around the ROI, and using only $p_{\text{good}}$ for scoring. \\
    \texttt{dl\_stack}~\citep{legorreta2020DL} & (1) & Stack-wise deep learning-based quality assessment. \\
    \midrule
    \multicolumn{3}{l}{\textsc{Metadata-based metrics}}\\
    \midrule
    \texttt{im\_size}  & (5)& voxel size (in-plane $x$ and $y$ and through-plane), as well as voxel-size and in-plane pixel dimension.\\
    \midrule
    \end{tabular}}
    
    \label{tab:metrics}
\end{table}

\newpage
\FloatBarrier
\subsection{Quality rating protocol}
In this section, we describe and illustrate the protocol that we used for quality rating in the study. The quality rating is defined in the context of downstream super-resolution reconstruction (SRR) from multiple low-resolution acquisitions in different orientations (axial, coronal, sagittal). The quality assessment specifically aims at quantify how suitable a given raw T2 weighted volume is for SRR, and is \textit{not} a radiological assessment of the image.

\textbf{Rating protocol}
Using a FetMRQC report, our quality rating was consisted of going through the following questions.
\begin{itemize}[noitemsep,topsep=0pt]
    \item \textbf{Motion-related artifacts.} Fetal motion can induce in-plane and through-plane artifacts.
    \begin{itemize}[noitemsep,topsep=0pt]
        \item Is there in-plane motion: signal drop/void, blurring, aliasing, ringing artifacts?
        \item Is there through-plane motion: loss of structural continuity in neighbouring slices~\citep{gholipour2014fetal,uus2022retrospective}?.
    \end{itemize}
    \item \textbf{Bias related artifacts.} Bias field is typically described as a smoothly varying spatial inhomogeneity that alters image intensities that otherwise would be constant for the same tissue type regardless of its position in the image ~\citep{vovk2007review}. In practice, it often appears like a shade on a part of an image, and is visible both in-plane and through-plane. 
    \begin{itemize}[noitemsep,topsep=0pt]
        \item Is there in-plane bias: differences in intensity within a single tissue on a given slice?
        \item Is there through-plane bias: difference in intensity within a single tissue across slices?
    \end{itemize}
    \item \textbf{Miscellanous.}
    \begin{itemize}[noitemsep,topsep=0pt]
        \item Is the image paritcularly noisy: grainy appearance?
        \item Is the brain entirely contained on the image?
    \end{itemize}
\end{itemize}
After answering the questions, the rater is asked to provide a score for global quality. Examples below show the scores that we assigned to various images.

\subsubsection{Example cases}
\begin{figure*}[!ht]
    \centering
    \includegraphics[width=.42\linewidth]{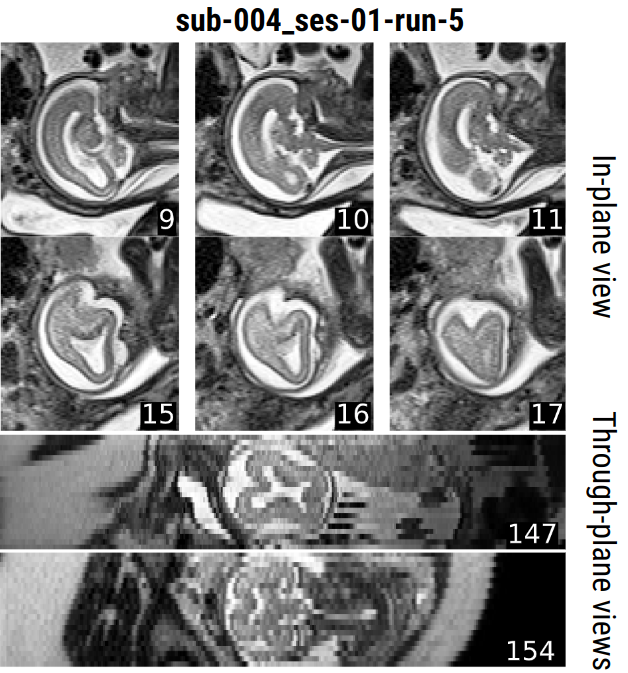}\hfill
    \includegraphics[width=.42\linewidth]{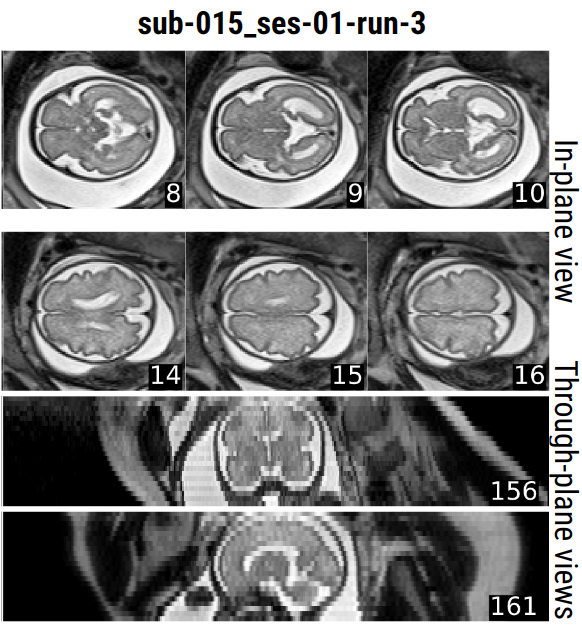}
    \caption{\textbf{Excellent quality images}. (Left) \textsc{Acceptable} to \textsc{excellent} quality sagittal image. No clear artifacts are visible in the in-plane view, and no large patterns of motion are view are present on the through-plane views. Moreover, the structure of the brain is clearly distinguished on all three planes. (Right) \textsc{Excellent} quality axial image. No artifacts are visible on any part of the image. The intensity of the tissues is spatially homogeneous, indicating a low bias, and no in-plane or through-plane affecting structural integrity are visible. } 
    \label{fig:excellent1}
    \includegraphics[width=.4\linewidth]{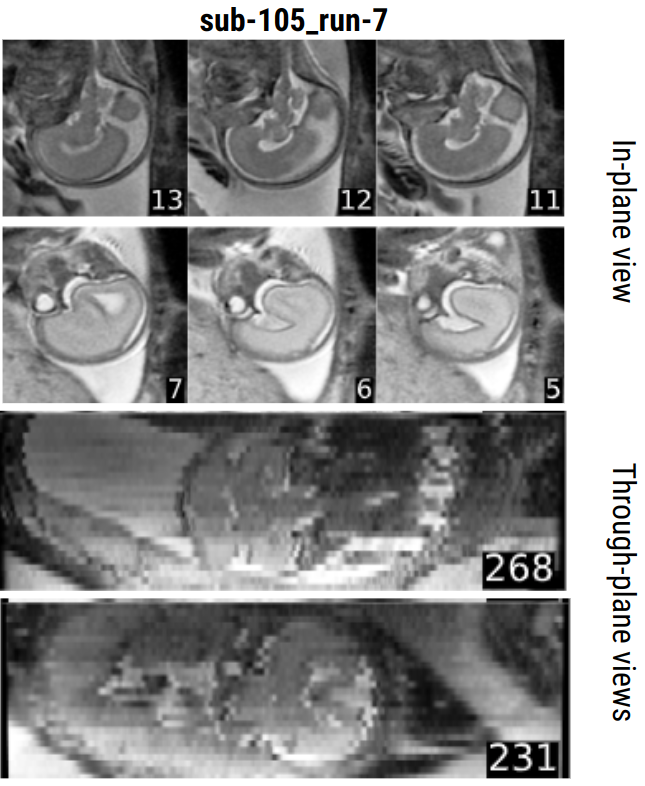}\hfill
    \includegraphics[width=.4\linewidth]{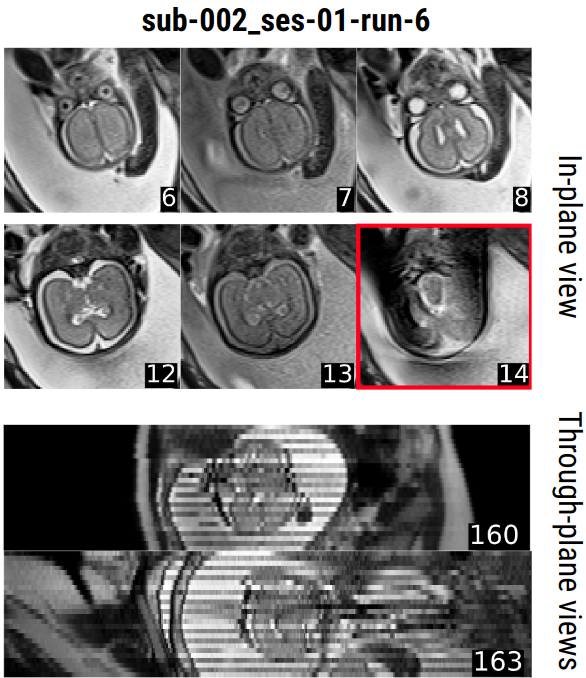}
    \caption{\textbf{Acceptable quality images.} \textbf{(Left)} \textsc{Acceptable} quality sagittal image. Although some ringing artifacts are visible outside the brain on slices 6 and 7, all brain structures are clearly visible in-plane. A strong bias field can be viewed between the top and bottom row of in-plane slices, as well as through-plane view. There is also moderate motion, viewed in  the through-plane view 268 where one sees various blocks of slices look disconnected from each other. \textbf{(Right)} \textsc{Poor} to \textsc{Acceptable} quality coronal image. On both in-plane and through-plane images, a clear intensity discontinuity is visible, suggesting a \textit{strong} bias bield. In addition, one sees on slice 14 a signal drop. No stair-like motion is visible, but a sharp discontinuity is seen on slice 160. This was rated as moderate motion.} \label{fig:acceptable}
    \end{figure*}

\begin{figure*}[!ht]
    \centering
    \includegraphics[width=.4\linewidth]{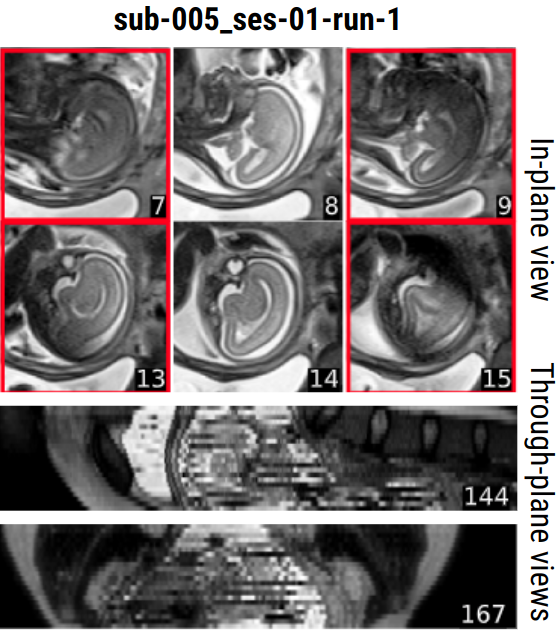}
    \hfill
    \includegraphics[width=.4\linewidth]{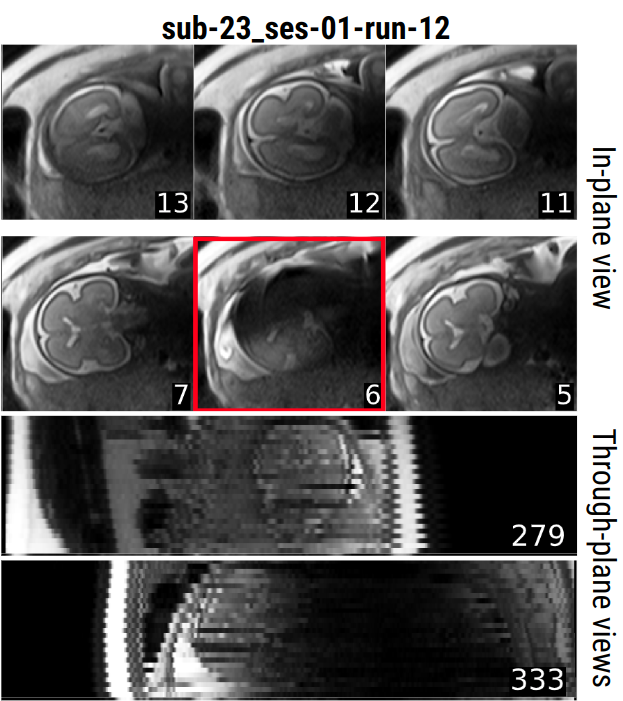}
    \caption{\textbf{Poor quality images.} \textbf{(Left)} \textsc{Poor}-to-\textsc{Exclude} quality sagittal image. Multiple slices are affected by signal drops suggesting heavy motion. Unsurprisingly, this leads to a poor structural integrity on the through-plane slices: it is very difficult to recognize the brain structure. \textbf{(Right)} Low \textsc{Poor} quality coronal image. A strong bias field is visible on all in-plane slices (the bottom of each slice is much darker than the top for the same tissue), and also through-plane: on 279, the left is darker than the right, and on 333, the left is clearer than the right. In addition, a typical "staircase" motion is viewed through-plane, as well as a signal drop on slice 6.} \label{fig:poor}
\end{figure*}

We now review six cases from our data, featuring different gestational ages and orientations with explanation of the artifacts in the captions. This review does not aim to be exhaustive. Figure \ref{fig:excellent1} shows two cases of \textsc{excellent} quality acquisitions. The other images show typical artifact patterns that can be found in fetal images. Figure \ref{fig:acceptable} shows acceptable and \textsc{poor}-to-\textsc{acceptable} quality images. Figure \ref{fig:poor} shows poor quality images.


\newpage
\FloatBarrier
\subsection{FetMRQC ablation studies}\label{sec:ablations}
Various components such as IQM standardization, feature selection, dimensionality reduction can be included in FetMRQC. We used nested cross-validation to automatically perform model selection and evaluation without introducing optimistic biases~\citep{varoquaux2017assessing}. We then had three variants of FetMRQC to compare:
\begin{itemize}[noitemsep]
    \item \textbf{Vanilla} FetMRQC: No preprocessing, use all IQMs and fit a random forest.
    \item \textbf{Nested CV} FetMRQC: Preprocessing, feature selection and various possible models, with selection performed using nested CV. The parameters are the one of Table \ref{tab:params} below.
    \item \textbf{ComBat + Nested CV} FetMRQC: ComBat~\citep{johnson2007adjusting}, preprocessing, feature selection and various possible models, with selection performed using nested CV. The parameters are the one of Table \ref{tab:params}.
\end{itemize}

\begin{figure}[t]
            \centering
        \includegraphics[width=\linewidth]{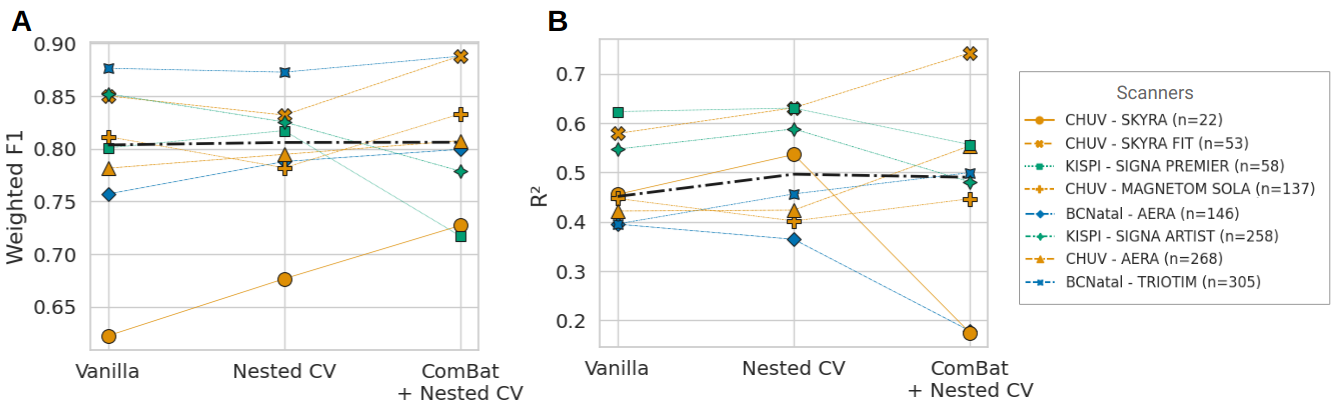}
        \caption{\textbf{Scanner-wise results for the QA/QC ablation study.} This is the breakdown of Table~\ref{tab:ablation}. \textbf{A} -- Weighted F1 score for the QC task, for each scanner used in LoSo CV. \textbf{B} -- $\text{R}^2$ for the QA task for each scanner used in LoSo CV.}
        \label{fig:ablations}
    \end{figure}

We evaluated each of these methods in a leave-one-scanner-out (nested) CV and report the results on Table \ref{tab:ablation}. The differences between Vanilla FetMRQC and the variants were tested with Welch's t-test (to take into account the unequal variance across samples), but none of the differences were found to be statistically significant. The breakdown of the results is shown on Figure~\ref{fig:ablations}, where we see that no method manages to provide an improvement for all scanners. While some scanners get a better performance, the performance is also decreases for other scanners. This is most clearly seen for Combat + Nested CV in the QA task (Figure~\ref{fig:ablations}B).

The full nested cross-validation very largely increases the computational time required to train the model. Given the IQMs, vanilla FetMRQC takes around 5 to 10 seconds to be trained. Nested CV evaluates 1004 models (regression) and 1344 models (classification), and parts like the Winnow algorithm make the overall training slower. Our simple implementation, using 5 parallel workers, took around one day to run. While this could certainly be greatly improved, it is clear that nested CV brings a much larger computational burden compared to vanilla FetMRQC. In this case, as it did not bring any significant benefit, we chose to only rely on the vanilla version of FetMRQC: using all IQMs without scaling and with a random forest.   

\review{While these ablation studies focus on various pre-processing steps and using different models, we also carried out additional ablation studies where each of the regression or classification model were trained using different parameters (e.g. larger or smaller forests, different fitting criteria, regularization, etc.). We used a random grid of parameters in each nested CV fold, and had to disable the Winnow algorithm for the training time to be reasonable. The results of this experiment (not presented) were largely similar to the ablation below.}

\begin{table}[h]
    \centering
    \captionof{table}{Parameters automatically optimized by the inner loop of the nested CV.}
    \label{tab:params}
    \begin{tabular}{lc}
    \toprule
    Model step & Parameters\\
    \midrule
    Remove correlated features & $\text{Threshold} \in \{0.8,0.9\}$; Disabled  \\
    Data Scaling & \makecell{Standard (group) scaling, Robust (group) scaling,\\ Quantile (group) scaling No scaling}  \\
    Winnow algorithm & Enabled, Disabled\\
    PCA & Enabled, Disabled\\
    Regression models & \makecell{Linear regression, Gradient\\ boosting, Random Forest} \\
    Classification models & \makecell{Logistic regression, Random Forest,\\ Gradient Boosting, AdaBoost}\\
    \bottomrule
    \end{tabular}
\end{table}
\begin{table}[t]
\centering
\captionof{table}{\textbf{Quality control and assessment ablation study.} LoSo CV was performed using vanilla FetMRQC, as well as nested cross-validation for hyperparameter tuning. A third variant preprocessed the data using ComBat~\citep{johnson2007adjusting} prior to performing nested CV. Results are the median cross-validation performance. The number in parentheses is the average worst-performing cross-validation fold. }\label{tab:ablation}
    \vspace{0.5cm}
    \centering
     \resizebox{\linewidth}{!}{
    \begin{tabular}{lcccclccc}
    \toprule	
    \multicolumn{5}{l}{\textsc{Quality control (classification)}} & \multicolumn{4}{l}{\textsc{Quality assessment (regression)}}\\
    \cmidrule(l){1-5} \cmidrule(l){6-9}
     & Weighted F1 ($\uparrow$)  & ROC AUC ($\uparrow$) & Precision ($\uparrow$) & Recall ($\uparrow$) 
    & & $R^2$ ($\uparrow$)  & Spearman ($\uparrow$) & MAE ($\downarrow$) \\
    \midrule
    \multicolumn{5}{c}{Leave-one-Scanner-out cross-validation} & \multicolumn{4}{c}{Leave-one-Scanner-out cross-validation}\\ 
    \midrule
    Nested CV           & 0.81 (0.68)  & 0.79 (0.64) & 0.85 (0.73) & 0.83 (0.79) & Nested CV           & 0.50 (0.36)  & 0.73 (0.70) & 0.55 (0.54) \\
    ComBat+Nested CV  & 0.80 (0.73)  & 0.79 (0.71) & 0.86 (0.71) & 0.89 (0.79)  &ComBat+Nested CV  & 0.49 (0.18)  & 0.75 (0.68) & 0.57 (0.60)\\
    Vanilla             & 0.81 (0.62)  & 0.78 (0.58) & 0.89 (0.69) & 0.83 (0.79)  &Vanilla             & 0.45 (0.39)  & 0.74 (0.71) & 0.56 (0.51)\\
    \bottomrule
    \end{tabular}}
    
\end{table}

\end{document}